\documentclass[fleqn,usenatbib]{mnras}

\usepackage{newtxtext,newtxmath}
\usepackage{gensymb}
\usepackage{amsmath}
\usepackage{threeparttable}
\usepackage{xspace}
\usepackage{mathtools}

\def\gaiam{\textit{Gaia}\xspace}
\def\gaia{\textit{Gaia}~DR3\xspace}
\def\spitzer{\textit{Spitzer}\xspace}
\def\massage{\texttt{MassAge}\xspace}
\def\sptclass{\texttt{SPTCLASS}\xspace}
\def\msun{$\mathrm{M}_{\odot}$\xspace}
\def\teff{$\mathrm{T}_{\mathrm{eff}}$\xspace}
\def\logLs{$\log\mathrm{L}_\ast$\xspace}
\def\vsini{$v\sin i$\xspace}
\def\logg{$\log g$\xspace}
\def\Halpha{H$\alpha$\xspace}
\def\mstar{M$_\ast$\xspace}
\def\rstar{R$_\ast$\xspace}
\def\yrinv{yr$^{-1}$\xspace}
\def\dotm{$\dot{\mathrm{M}}$\xspace}

\def\av{$\mathrm{A_V}$\xspace}
\def\al{$\mathrm{A_{\lambda}/A_V}$\xspace}

\def\logL{\mathrm{log(L/L_{\odot})}}

\usepackage[T1]{fontenc}

\DeclareRobustCommand{\VAN}[3]{#2}
\let\VANthebibliography\thebibliography
\def\thebibliography{\DeclareRobustCommand{\VAN}[3]{##3}\VANthebibliography}

\usepackage{graphicx}	
\usepackage{amsmath}	
\usepackage[normalem]{ulem}

\title[A HECTOSPEC view of the ONC]{A Hectospec Spectroscopic Census of Kinematic Members in the Orion Nebula Cluster}

\author[J. Hernández et al.]{Jesús Hernández$^{1}$\thanks{E-mail: hernandj@astro.unam.mx},
Ezequiel Manzo-Martínez$^{1,2}$,
Lucía Adame$^{1}$,
Sergio Sánchez-Sanjuán$^{1}$,
Nuria Calvet$^{2}$,
\newauthor
David Fernandez$^{3}$,
Cesar Briceno$^{4}$,
Javier Serna$^{5}$,
Giovanni Pinzón$^{6}$,
Lucia Martinez$^{1}$,
Tom Megeath$^{7}$,
\newauthor
John J. Tobin$^{8}$,
Carlos Román-Zuñiga$^{1}$,
Karina Maucó$^{1}$,
Corina Zepeda$^{1}$,
Andrea Bonilla-Barroso$^{9}$,
\newauthor
Maria Gracia Batista$^{9}$,
Javier Ballesteros-Paredes$^{10}$, 
Lori Allen$^{11}$,
Luisa Zamudio-Ruvalcaba$^{12}$,
\newauthor
Fabian Ahumada$^{3}$,
Alejandro Alfonso$^{13}$,
Johana Contreras$^{6}$,
Jhordan Peña$^{6}$ 
\\
$^{1}$Instituto de Astronomía, Universidad Nacional Autónoma de México, Apartado Postal 106, C. P. 22800, Ensenada, B. C., México\\
$^{2}$Department of Astronomy, University of Michigan, 1085 South University Avenue, Ann Arbor, MI 48109, USA\\
$^{3}$Departamento de Astronom{\'i}a, Universidad de La Serena, 1700000 La Serena, Chile\\
$^{4}$SOAR Telescope/NSF NOIRLab, Casilla 603, La Serena, Chile\\
$^{5}$Homer L. Dodge Department of Physics and Astronomy, University of Oklahoma, Norman, OK 73019, USA\\
$^{6}$Observatorio Astron\'omico Nacional, Facultad de Ciencias, Universidad Nacional de Colombia, Bogot\'a, Colombia\\
$^{7}$ Ritter Astrophysical Research Center, Department of Physics and Astronomy, University of Toledo, Toledo OH, 43606, USA\\
$^{8}$National Radio Astronomy Observatory, 520 Edgemont Road, Charlottesville, VA 22903, USA\\
$^{9}$Universidad de los Andes, Departamento de F\'isica, Cra. 1 No. 18A-10, Bloque Ip, A.A. 4976 Bogot\'a, Colombia\\
$^{10}$Instituto de Radioastronom\'ia y Astrof\'isica, UNAM, campus Morelia. PO Box 3-72. 58090. Morelia, Michoac\'an, M\'exico\\
$^{11}$ NSF NOIRLab, 950 N. Cherry Ave., Tucson, AZ 85719, USA\\
$^{12}$ Department of Astronomy, Boston University, 725 Commonwealth Avenue, Boston, MA 02215, USA\\
$^{13}$ Universidad Pedagógica y Tecnológica de Colombia. Sede Central Tunja – Boyacá – Colombia\\
}

\date{Accepted for publication in MNRAS (16 October 2026}

\pubyear{\the\year{}}

\begin{document}
\label{firstpage}
\pagerange{\pageref{firstpage}--\pageref{lastpage}}
\maketitle

\begin{abstract}

We present a comprehensive spectroscopic census of the Orion Nebula Cluster (ONC), based primarily on low-resolution optical spectroscopy obtained with MMT/Hectospec. We derive spectral types, youth indicators, and accretion diagnostics for 1513 stars located in the ONC region, enabling the identification and characterization of pre--main-sequence stars. We combine these measurements with a kinematic membership analysis based on \gaiam Data Release 3 astrometry to identify and confirm cluster members. Using these combined diagnostics, we identify and characterize 780 bona fide T Tauri stars associated with the ONC.
In addition, we construct a census of protoplanetary discs using infrared photometry obtained with the IRAC and MIPS instruments aboard the \spitzer Space Telescope. We further derive stellar masses, ages, and extinctions using the \massage code, which combines photometric and spectroscopic inputs with stellar evolutionary models,
and employs a Monte Carlo approach for uncertainty propagation. The resulting homogeneous catalogue of ONC members provides spectroscopic characterizations, kinematic membership constraints, disc diagnostics, and self-consistent estimates of stellar physical parameters. This dataset offers a refined view of the stellar population of the ONC and provides a robust foundation for investigating its kinematic structure, age distribution, and protoplanetary disc population.
\end{abstract}




\section{Introduction}

The Orion Nebula Cluster (ONC) is one of the nearest and youngest regions of high-mass star formation,
with an age of $\sim$1-5 Myr and located at a distance of $\sim$400 pc \citep[e.g.,][]{Hillenbrand1997, Hillenbrand2013, Jeffries2011, Muench2008, Kounkel2018, Briceno2019}. Its proximity and high stellar density make it an ideal laboratory for studying the early stages of stellar evolution, including disc evolution, accretion processes, and the dynamical assembly of young clusters. Despite decades of study, fundamental questions regarding its star formation history, dynamical state, and age spread remain open \citep[e.g.,][]{Muench2008, Jeffries2011, Getman2014, Jerabkova2019, Kim2019, Fujii2022, Megeath2022, Theissen2022, Alzate2023, Wei2024}.

Optical spectroscopy provides a robust, direct diagnostic of pre-main-sequence (PMS) stellar populations by leveraging well-characterized spectral features sensitive to stellar youth. Low-mass PMS stars (\mstar $\lesssim$ 2 \msun), commonly known as T Tauri stars (TTS), are traditionally classified into classical T Tauri stars (CTTS) and weak-lined T Tauri stars (WTTS). 
CTTS are stars undergoing active accretion from their circumstellar discs, whereas WTTS show little or no evidence of accretion. Strong \Halpha emission is one of the primary diagnostics used to distinguish CTTS from WTTS, although it may also serve as an indicator of stellar chromospheric activity. In addition, the presence of the \ion{Li}{I} $\lambda6708$ \AA\ absorption line (hereafter \ion{Li}{I}) serves as a robust youth diagnostic in low-mass stars \citep{Jeffries2006, Soderblom2014, Briceno2019, Hernandez2023}.

Many spectroscopic surveys in the ONC have used these diagnostics to confirm membership \citep[e.g.,][]{Aurora2005, Gabor2008, Hillenbrand1997, Hillenbrand2013, Szegedi2013, Fang2017, Fang2021, Briceno2019, Saad2024, Sizemore2024, Piscarreta2025}. Radial velocities and proper motions have been extensively used to identify kinematically coherent members, unveiling complex dynamical structures and a close correspondence between the stellar and the gaseous components of the ONC \citep{Aurora2005, Gabor2008, Tobin2009, Kounkel2016, Kounkel2018, Jerabkova2019, Sergio2024, Wei2024}. 
In particular, recent studies have revealed multiple kinematically distinct substructures within the ONC. For example, \citet{Sergio2024} identified three stellar substructures whose members exhibit different proper-motion orientations, despite having largely indistinguishable parallax distributions. The origin of these substructures remains uncertain and may provide important clues about the star-formation history of the region.

Based on colour-magnitude diagrams,
\citet{Beccari2017} proposed that the stellar population of the ONC is composed of three distinct sequences corresponding to separate star-formation episodes. Supporting this scenario, \citet{Jerabkova2019} found evidence that the parallax and proper-motion distributions differ among the stars belonging to these photometric sequences. In contrast, \citet{Alzate2023}, accounting for the effects of unresolved binaries, 
found that the observed stellar population is consistent with a roughly constant star formation rate over the interval $\sim$0.3 and 5 Myr, rather than with a sequence of discrete star-formation episodes. 
These contrasting interpretations highlight that the star-formation history and dynamical assembly of the ONC remain open questions. They also emphasize the importance of combining spectroscopic diagnostics with precise astrometric information to robustly identify cluster members and disentangle overlapping stellar populations.

In addition, characterizing the circumstellar disc population in the ONC is fundamental to understanding the early evolutionary phases of protoplanetary discs. 
For example, \citet{Briceno2019} noted that \citet{Hillenbrand1998} estimated that $\sim$70\% of the young stellar population in the Orion Nebula Cluster are CTTS, indicative of a large fraction of disk-bearing stars, a result confirmed by
\citet{Haisch2001}.
Based on multi-epoch \spitzer IRAC and MIPS photometry, \citet{Megeath2012} identified a large population of disc-bearing stars and protostars, and showed that approximately half of the sources exhibit significant mid-infrared variability, likely driven by changes in accretion and inner disc structure. 
Subsequent studies \citep{Megeath2016, Manzo2020} have shown that, despite the young age of the ONC, a fraction of discs already exhibits signs of evolution, indicating a diversity of evolutionary stages within the cluster, spanning a wide range of accretion rates $(-10\lesssim \log(\dot{M}) \lesssim -8$, with \dotm in \msun\yrinv) and exhibiting a non-negligible degree of dust settling of large grains towards the midplane. 
However, a significant fraction of relatively evolved discs (e.g., transitional discs) may belong to an older foreground population along the line of sight toward the ONC \citep{DaRio2016}.

In this context, large and homogeneous spectroscopic surveys are essential to fully characterize the stellar content of the ONC. By combining spectral classifications, youth and accretion diagnostics, as well as precise kinematics from \gaiam Data Release 3 \citep[\gaia;][]{Gaia2023}, it is possible to construct a refined and self-consistent census of cluster members. Furthermore, placing these stars on the Hertzsprung–Russell diagram (HRD) enables the estimation of stellar masses and ages, providing key constraints on the star formation history and early evolution of the cluster. 

In this work, we present a large optical spectroscopic census of the ONC based mainly on Hectospec observations, comprising more than 1500 stars across the cluster field. We derived spectral types and assessed stellar youth and accretion through the analysis of \ion{Li}{I} and \Halpha emission, respectively.
We combine these data with astrometric measurements from \gaia to identify kinematically coherent members of the ONC and estimate stellar masses and ages. To complement our study, we also performed a protoplanetary disc census with \spitzer infrared photometry. 

This paper is organized as follows. Section~\ref{sec:spec} describes the spectroscopic data sets. Section~\ref{sec:members} presents the youth indicators and membership analysis. Section~\ref{section:Massage} derives stellar masses and ages. Section~\ref{sec:infrared} focuses on the infrared analysis, disc census and mass accretion rates estimates. Finally, Section~\ref{sec:Conclusions} presents the main conclusions.

\section{Spectroscopic database}
\label{sec:spec}

In this section, we describe the spectroscopic data set used in this work, including observations obtained with MMT/Hectospec and several complementary instruments, which together provide low-resolution optical spectra for 1513 stars located in the ONC region. This spectroscopic data set constitutes the main sample analysed throughout this work. 
The initial Hectospec target list was constructed primarily from the CIDA Variability Survey of Orion \citep{Briceno2005}, selecting variable stars with Two Micron All Sky Survey \citep[2MASS;][]{Skrutskie2006} counterparts that were located above the PMS locus assuming a distance of 400 pc, as well as sources with infrared excesses identified in a preliminary Spitzer catalogue used by \citet{Megeath2012}. These sources were assigned the highest priority in the Hectospec fiber-positioning algorithm. However, because of the constraints of the fiber positioner, not all high-priority targets could be observed. Subsequent observing campaigns using other spectroscopic facilities extended the target selection to the brighter population (V$<$15 mag), giving priority to sources with disks identified by \citet{Megeath2012}, as well as to sources with Gaia parallaxes and proper motions \citep{GAIA2016} consistent with membership in the Orion region. Along with the fact that spectroscopic observations are brightness limited,
the resulting spectroscopic sample is not a complete or homogeneous census of the underlying stellar population.

Some targets were observed more than once, either with the same or with different spectroscopic facilities, as indicated in the “Inst.” column of Table \ref{tab:main}. The procedure adopted to combine these multiple observations is described in Section \ref{sec:sptclass}.

\subsection{Hectospec - MMTO}
\label{sec:hsp_spec}
We obtained seven fields with the fiber-fed multi-object Hectospec (HSP) instrument mounted on the 6.5 m Telescope at the MMT Observatory (MMTO). 
HSP has 300 fibers that can be placed within a 1{\degree} diameter field of view. Each fiber subtends 1.5{\arcsec} on the sky \citep{Fabricant2005}. 
The observations were taken with a 270 lines mm$^{-1}$ grating, providing a spectral resolution of $\sim$6 {\AA} and spectral coverage of 3650–9200 \AA. All seven HSP fields were observed during the nights of 2009 February 02 and 04.

\begin{figure}
    \centering
    \includegraphics[width=0.9\columnwidth]{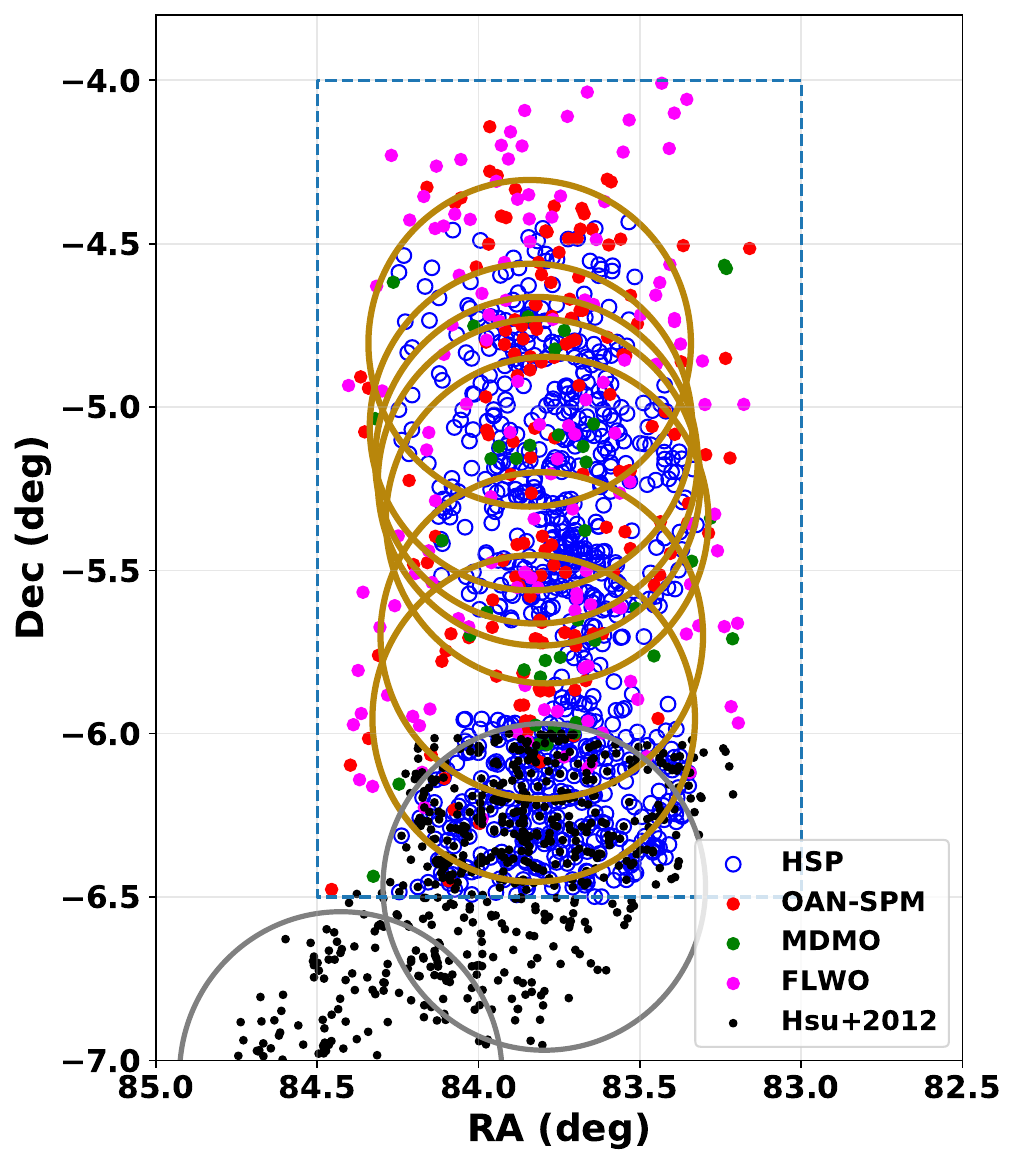}
    \caption{Spatial distribution of the stars analysed in this work. Symbols and colours represent spectra obtained with different spectrographs. Blue circles denote stars with Hectospec spectra obtained in this work, while black dots correspond to Hectospec observations from \citet{Tina2012}. Red, green, and magenta dots represent stars observed with spectrographs at the SPMO, MDMO, and FLWO observatories, respectively. The large dashed rectangle marks the studied region that includes the ONC, and the large open circles indicate the Hectospec fields analysed in this work (yellow) and those from \citet[][; gray]{Tina2012}, who studied the low-mass stellar population in L1641.}
    \label{fig:HSP}
\end{figure}

The HSP data were primary reduced using the standard HSP pipeline \citep{Mink2007}. Only the sky background subtraction was performed using an alternative approach.
The standard pipeline for this instrument assumes that the sky background does not vary significantly across the field and therefore uses a single sky spectrum constructed by combining the spectra from all sky fibers within the HSP field. However, because the sky background across the ONC is highly variable, we performed the sky subtraction step using an independent off-field sky observation obtained 5{\arcmin} apart from each target field. For each HSP field, we obtained three 900 s exposures for the target field and one 900 s exposure for the off-field sky, taken just before or after the corresponding target-field observations.
A total of 892 stars have at least one spectrum with a signal-to-noise ratio (S/N) greater than 30 in these fields. Figure~\ref{fig:HSP} shows the location of the seven HSP fields and the region studied in this work (hereafter the ONC region; 83{\degree} $<$ RA $<$ 84.5{\degree} and $-$6.5{\degree} $<$ Dec $<$ $-$4.0{\degree}). 
The selected spatial extent was motivated by the distributions of ONC members reported in previous studies \citep[e.g.][]{Muench2008, Peterson2008, Megeath2016}.
We also included 228 stars analysed by \citet{Tina2012} as part of their spectroscopic census of the stellar population in L1641 that fall in the ONC region. 
These observations were obtained with the same Hectospec configuration and observing strategy as those described above.
Thus, we have a total of 1120 stars with HSP data in the ONC region.\\

\subsection{Boller and Chivens - OAN-SPM} \label{sec:SPMO_spec}
Low-resolution, long-slit spectra were obtained with the Boller \& Chivens spectrograph mounted on the 2.1 m telescope at the San Pedro M\'artir Observatory (OAN-SPM) during several observing campaigns: November 2015, January 2016, November/December 2016, January 2017, and February 2020. The last campaign was designed to observe the bright kinematic member candidates (V $<$ 14) selected using proper motions and parallaxes from \gaiam DR2 \citep[e.g., ][]{Kounkel2018} that had not been observed previously. We used a 400 lines mm$^{-1}$ dispersion grating along with a 2{\arcsec} slit width, providing a spectral resolution of $\sim$ 6 {\AA} with a spectral coverage of $\sim$ 3300 {\AA} centered at $\sim$ 5500 \AA. The observations were reduced and calibrated following standard procedures using the IRAF packages \texttt{twodspec} and \texttt{onedspec}. These campaigns provided spectra for 192 stars located in the ONC region.\\

\subsection{Boller and Chivens - MDMO} \label{sec:MDMO_spec}

Low-resolution, long-slit spectra were obtained with the OSU (Ohio State University) Boller \& Chivens CCD spectrograph (CCDS) mounted on the 1.3 m McGraw-Hill telescope at the MDM Observatory (MDMO) during several observing campaigns: December 2008, November 2009, March 2011, December 2011, December 2013, and November 2014. We used the 158 lines mm$^{-1}$ dispersion grating with a 1{\arcsec} slit width, providing a spectral resolution of $\sim$6.5 {\AA}\footnote{The nominal resolution of 4.1 {\AA} provides spectra undersampled at this telescope} with a spectral coverage of $\sim$3400 {\AA} centered at $\sim$5300 {\AA}. As in Section~\ref{sec:SPMO_spec}, the reduction and calibration were performed using standard IRAF procedures. We obtained spectra of 47 stars in the ONC region with this instrument.\\

\subsection{FAST - FLWO} \label{sec:FAST_spec}

We collected spectra for 222 stars in the ONC region, observed with the FAST spectrograph mounted on the 1.5 m telescope at the Fred Lawrence Whipple Observatory \citep[FLWO; ][]{Fabricant1998}, from the FAST Public Archive\footnote{\url{http://tdc-www.harvard.edu/cgi-bin/arc/fsearch}}. 
These data were obtained through the FAST programs 112: Orion PMS Candidates \citep[e.g.,][]{Briceno2005,Briceno2007} and 89: Ae/Be Stars in OB Associations \citep[e.g.,][]{Hernandez2014}, which used the FAST COMBO standard configuration with a 300 lines mm$^{-1}$ dispersion grating and a 3\arcsec wide slit. This combination provides a spectral coverage of $\sim$3400 {\AA} centered at $\sim$5500 \AA, with a spectral resolution of $\sim$6 \AA. The spectra were reduced at the Harvard–Smithsonian Center for Astrophysics using software developed specifically for FAST COMBO observations.\\

\section{Memberships and youth features}
\label{sec:members}

\subsection{Spectroscopic analysis}
\label{sec:sptclass}

We obtained spectral types and equivalent widths (EWs) of \Halpha and \ion{Li}{I} $\lambda6708$ {\AA} for the main sample described in 
Section~\ref{sec:spec} using the \sptclass code \citep{Hernandez2004,Hernandez2017}, a semi-automatic spectral analysis tool written in IRAF/IDL, which employs empirical relations between EWs of selected atomic and molecular absorption features and spectral types to classify and characterize young stellar objects \citep[YSOs;][]{Hernandez2004,Hernandez2014, Hernandez2023, Aurora2005,Tina2012, Briceno2019}.
For each spectral feature, the EW is obtained by measuring the decrease in flux relative to the local continuum, estimated by interpolating between two adjacent bands. 
The spectral features measured by this procedure are largely insensitive to reddening and to the S/N of the spectra (provided that the S/N is sufficient to detect the feature). Moreover, spectral types determined with \sptclass are largely independent of luminosity because most of the selected indices are primarily sensitive to effective temperature (\teff) and only weakly dependent on the surface gravity (\logg) of the star \citep{Hernandez2004}. Table~\ref{tab:main} lists the results of the spectroscopic analysis with \sptclass.\\

It is important to note that \sptclass assumes that the spectrum originates from a single star and does not account for the effects of hot continuum emission produced by accretion shocks (i.e., veiling) or other non-photospheric contributions, such as infrared excess emission arising from the inner circumstellar disc \citep[e.g.,][]{Manara2021, Perez2024}. 
In veiled stars, spectral features appear weaker and, for K- and M-type stars, the spectral types derived with \sptclass may be biased toward earlier types. 
Consequently, the inferred \teff may be systematically higher than the actual effective temperature and should therefore be regarded with caution for veiled stars\citep[i.e.,][]{Tina2012}.
Furthermore, \sptclass does not account for thermal inhomogeneities on the stellar surface, such as cool stellar spots. Because the contrast between spots and the surrounding photosphere is larger at optical wavelengths than in the near-infrared, \teff derived from optical spectra may be systematically higher than those obtained from near-infrared spectroscopy \citep[e.g.,][]{Flores2022, Gangi2022, Perez2024}. 
These biases may propagate into the derived stellar masses, ages, and accretion rates, particularly for strongly veiled or heavily spotted objects.
Accounting explicitly for thermal surface inhomogeneities, unresolved binarity, or accretion-related veiling requires more sophisticated spectral modelling approaches that incorporate additional free parameters describing the physical properties of these phenomena. Although such methods may provide more accurate estimates for individual objects, they also introduce significant parameter degeneracies, particularly when applied to low-resolution spectra. In addition, these effects are often time-dependent, causing the observed spectrum to vary with stellar rotational phase, binary orbital configuration, or changes in the accretion rate. A detailed treatment of these effects is therefore beyond the scope of this work.\\

We applied \sptclass to the main sample. 
Some stars were observed multiple times, either with the same spectrograph or with different instruments. In these cases, each spectrum was analyzed independently with SPTCLASS, and the final spectral types and EWs reported in Table \ref{tab:main} were generally computed as weighted averages of the individual measurements.
We successfully classified 1,243 stars as K- or M-type (hereafter KM-type), 161 stars with spectral types between F5 and K0 (hereafter solar-type), and 109 stars with spectral types earlier than F5 (hereafter early-type). Using \sptclass in interactive mode, we visually inspected all low-resolution spectra and measured \Halpha and \ion{Li}{I} EWs by fitting Gaussian profiles, 
selecting interactively two points on either side of each spectral feature to define the local continuum. This interactive procedure also allowed us to assess the significance of the \ion{Li}{I} feature above the noise and the quality of the Gaussian fit to the line (flag\_\ion{Li}{I} in Table \ref{tab:main}).

SPTCLASS does not provide formal uncertainties for these EW measurements. Given the spectral resolution, the \ion{Li}{I} measurements are used primarily to assess the presence of the line, while the \Halpha equivalent widths are used to identify accreting stars and distinguish CTTS from WTTS. The latter classification is based on EW thresholds that are sufficiently separated that measurement uncertainties are generally not expected to affect the classification.
\\

\begin{table}
\centering
\caption{Main Sample}
\begin{tabular}{ll}
\hline
Column label & Description       \\
\hline \hline
2MASS\_ID   &  2MASS source name \\
Gaia DR3\_ID   &  \gaia source name \\
SpT   &  Spectral type \\
SpT\_err& Error in Spectral Type  \\
EW\_LiI   &  Equivalent Width of \ion{Li}{I} \\
flag\_LiI   &  flag for \ion{Li}{I} detection \\
EW\_\Halpha   &  Equivalent Width of \Halpha \\
Inst.  &  Spectral database  \\
Comments   &  notes about spectrum  \\
CB\_flag  & Acretion status from \citet{Briceno2019}\\
spec\_flag   & flag from spectral analysis (see Table~\ref{tab:classification_flags})\\
k\_flag   & flag for kinematic membership  \\
Big\_Cluster   & Clusters-big from \citet{Sergio2024}   \\
Small\_Cluster   & Clusters-small from \citet{Sergio2024}\\
\teff & median effective temperature\\
\teff low & lower limit of the effective temperature \\
\teff upp & upper limit of the effective temperature \\
{\av} & median visual extinction\\
{\av} low & lower limit of the visual extinction \\
{\av} upp & upper limit of the visual extinctions \\
$\logL$ & median stellar luminosity\\
{$\logL$} low & lower limit of the stellar luminosity \\
{$\logL$} upp & upper limit of the stellar luminosity \\
$[3.6]$ & \spitzer photometry, 3.6 \micron \\
e\_$[3.6]$ & error of the \spitzer photometry, 3.6 \micron \\
$[4.5]$ & \spitzer photometry, 4.5 \micron \\
e\_$[4.5]$ & error of the \spitzer photometry, 4.5 \micron \\
$[5.8]$ & \spitzer photometry, 5.8 \micron \\
e\_$[5.8]$ & error of the \spitzer photometry, 5.8 \micron \\
$[8.0]$ & \spitzer photometry, 8.0 \micron \\
e\_$[8.0]$ & error of the \spitzer photometry, 8.0 \micron \\
$[24]$ & \spitzer photometry, 24.0 \micron \\
e\_$[24]$ & error of the \spitzer photometry, 24.0 \micron \\
Cl\_M2012 & classification from \citet{Megeath2012} \\
flag\_3.6\micron & Infrared analysis: CIII, CII, CI \\
flag\_4.5\micron & Infrared analysis: CIII, CII, CI \\
flag\_5.8\micron & Infrared analysis: CIII, EV, CII, CI \\
flag\_8.0\micron & Infrared analysis: CIII, EV, CII, CI \\
flag\_24\micron & Infrared analysis: CIII, EV, CII, CI \\
log(L$_{acc}/L_{\odot}$)\_1 & accretion luminosity \\
log(L$_{acc}/L_{\odot}$)\_low\_1 & lower limit of the accretion luminosity \\
log(L$_{acc}/L_{\odot}$)\_upp\_1 & upper limit of the accretion luminosity \\
$\dot{M}$\_1 & mass accretion rate, \msun\yrinv \\
$\dot{M}$\_low\_1 & lower limit for the mass accretion rate, \msun\yrinv \\
$\dot{M}$\_upp\_1 & upper limit for the mass accretion rate, \msun\yrinv \\
log(L$_{acc}/L_{\odot}$)\_2 & accretion luminosity \\
log(L$_{acc}/L_{\odot}$)\_low\_2 & lower limit of the accretion luminosity \\
log(L$_{acc}/L_{\odot}$)\_upp\_2 & upper limit of the accretion luminosity \\
$\dot{M}$\_2 & mass accretion rate, \msun\yrinv \\
$\dot{M}$\_low\_2 & lower limit for the mass accretion rate, \msun\yrinv \\
$\dot{M}$\_upp\_2 & upper limit for the mass accretion rate, \msun\yrinv \\
\hline
\end{tabular}
\label{tab:main}
\begin{tablenotes}
\item Note: The subscripts 1 and 2 in the last columns identify quantities derived using the empirical accretion calibrations of Ingleby et al. (2013) and Alcalá et al. (2014), respectively.
\end{tablenotes}
\end{table}

\begin{table}
    \centering
    \caption{Classification criteria for the \texttt{spec\_flag} parameter (Table~\ref{tab:main}) based on the \ion{Li}{I} spectral-feature flag (\texttt{flag\_LiI}) and the accretion-status flag (\texttt{CB\_flag}). }   \label{tab:classification_flags}
    \begin{tabular}{lcc}
        \hline
        \hline
        spec\_flag & flag\_LiI & CB\_flag \\
        \hline
        CTTS        & ---        & \texttt{C} \\
        WTTS        & \texttt{Y} & \texttt{W} \\
        CWTTS       & \texttt{Y} & \texttt{WC} \\
        WTTSq       & \texttt{Q} & \texttt{W} \\
        CWTTSq      & \texttt{Q} & \texttt{WC} \\
        TTS         & \texttt{Y} & Undefined \\
        TTSq        & \texttt{Q} & Undefined \\
        Field\_stars & \texttt{N} & $\neq$ \texttt{C} \\
        \hline
    \end{tabular}
\end{table}

\subsubsection{KM-type stars}

Since lithium in low-mass stars can be depleted by 1–2 dex during the first 100 Myr \citep{Soderblom2014}, we used the presence of \ion{Li}{I} in absorption as an unambiguous spectroscopic indicator of youth in KM-type stars. Through visual inspection, we assigned a flag for the presence of \ion{Li}{I} in absorption in each spectrum: \texttt{Y} indicates a clear detection of the \ion{Li}{I} line, \texttt{N} indicates no detectable \ion{Li}{I}, and \texttt{Q} denotes a doubtful detection. These flags are listed in column flag\_LiI in Table~\ref{tab:main}\\

\begin{figure}
    \centering
    \includegraphics[width=0.9\columnwidth]{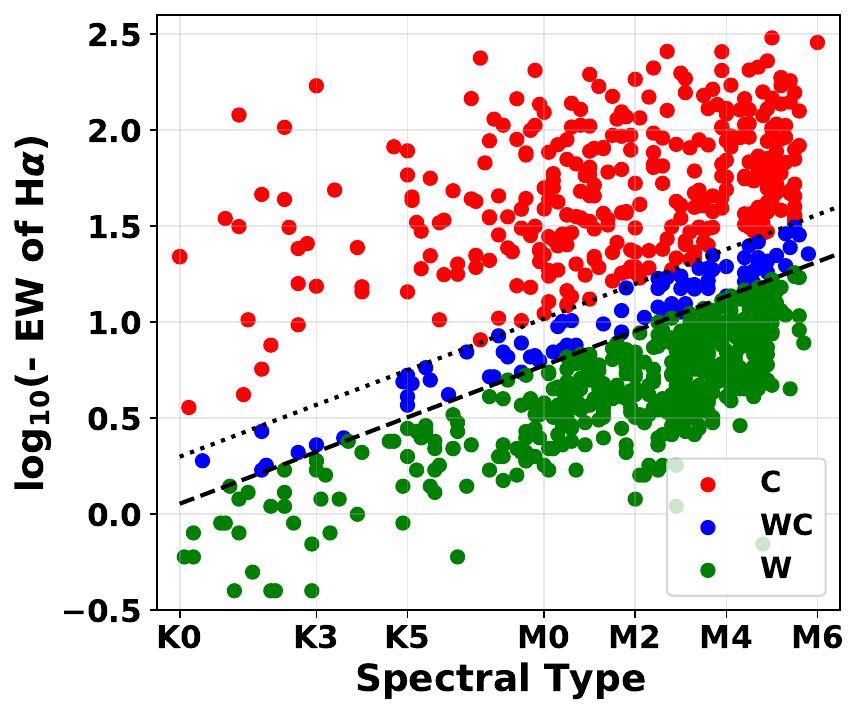}
    \caption{
    Logarithm of the \Halpha equivalent width as a function of spectral type. We identify 341 accretors (red), 423 non-accretors (green) and  112 WC stars (blue). The dotted and dashed lines indicate the classification boundaries defined by \citet{Briceno2019} for TTS.
    }
    \label{fig:CW_class}
\end{figure}

We also applied the criteria defined by \citet{Briceno2019} to estimate the accretion status of KM-type stars, based on the relation between the equivalent width of the \Halpha line and spectral type (see Figure~\ref{fig:CW_class}). In Table~\ref{tab:main}, we include an accretion status flag (CB\_flag), where \texttt{C} indicates accretors, \texttt{W} non-accretors, and \texttt{WC} stars likely in a transitional evolutionary stage approaching the end of their accretion phase \citep{Briceno2019}. Since the ONC region has a high stellar density and a highly variable sky background, some spectra may be affected by residual sky contamination. By visual inspection, we identified 153 KM-type stars showing apparent sky background residuals. For these objects, we did not measure the EW of \Halpha and have an undefined CB\_flag.\\

Combining the \ion{Li}{I} and $H\alpha$ classifications, we divided the KM-type stars into eight categories, identified by the \texttt{spec\_flag} column in Table~\ref{tab:main} and described in detail in Table~\ref{tab:classification_flags}. Since the strong non-photospheric continuum in highly accreting CTTS can significantly veil the \ion{Li}{I} line, we also classified as CTTS those stars showing marginal (68 objects) or no detectable (39 objects) \ion{Li}{I}. Including these sources, the final CTTS sample comprises 341 objects.

\subsubsection{Solar-type stars}

Since PMS stars earlier than K0 lack deep convective envelopes, solar-type stars do not significantly deplete lithium during their PMS evolution, making \ion{Li}{I} absorption an unreliable indicator of stellar youth in this spectral-type range \citep{Briceno2019,Soderblom2014}.
Consequently, the presence of lithium in absorption in solar-type stars primarily indicates that these objects are not evolved field stars belonging to the old disc population or post–main-sequence phases \citep{Briceno1997, Hernandez2004, Soderblom2014}. Additional youth indicators are therefore required to assess the evolutionary status of stars with spectral types earlier than K0.\\

In this spectral type regime, \Halpha emission provides a diagnostic of ongoing accretion and disc presence and is commonly used to identify the intermediate-mass counterparts of CTTS, known as Intermediate-Mass T Tauri Stars (IMTTS). 
We therefore primarily relied on the EW of \Halpha as a proxy for accretion activity in the solar-type sample.
We identified 40 IMTTS, of which 33 also exhibit a reliable visual detection of \ion{Li}{I}. 

Additionally, we classified 118 solar-type stars as non-accreting (spec\_flag = solar-type\_noacc), based on the presence of \Halpha in absorption, most of which show no detectable \ion{Li}{I}. Finally, seven stars lack \Halpha measurements due to background-subtraction issues and are labeled as solar-type\_undefined.

\subsubsection{Early-type stars}

Following the same approach adopted for solar-type stars, we used the EW of \Halpha to classify stars with spectral types earlier than F5. We identified 13 stars exhibiting \Halpha in emission, as expected for Herbig Ae/Be stars (HAeBe), which host actively accreting circumstellar discs \citep[e.g.,][]{Waters1998,Hernandez2004,Vioque2018}. These stars have been previously classified as HAeBe stars by \citet{Chen2016} or confirmed as members of the ONC by \citet{Hillenbrand2013}. \\

We did not detect \Halpha emission in 91 early-type stars, suggesting that accretion has either ceased or is occurring at levels below our detection threshold, as is typical of relatively evolved stellar populations (spec\_flag = early-type\_noacc). Additional membership diagnostics are therefore required to assess the nature of these objects, such as kinematic properties (Section~\ref{sec:kinmem}), their position in the HRD (Section~\ref{section:Massage}), or infrared excess emission (Section~\ref{sec:infrared}).

It is important to note that the \Halpha-based classifications adopted for solar-type and early-type stars should be regarded as proxies for accretion activity. Unlike KM-type stars, whose photospheres generally do not exhibit intrinsic \Halpha absorption, the photospheres of solar-type and earlier-type stars naturally display \Halpha absorption features. As a result, stars classified as non-accreting based on a positive \Halpha EW may still harbour a weak emission component partially filling the underlying photospheric absorption profile. A more robust assessment of accretion activity would require subtracting the photospheric contribution to the \Halpha line, which is beyond the scope of this work.

\subsubsection{Comparison with other spectral type classifications}
\label{s:spt_comparison}

\begin{figure*}
    \centering   \includegraphics[width=0.8\textwidth]{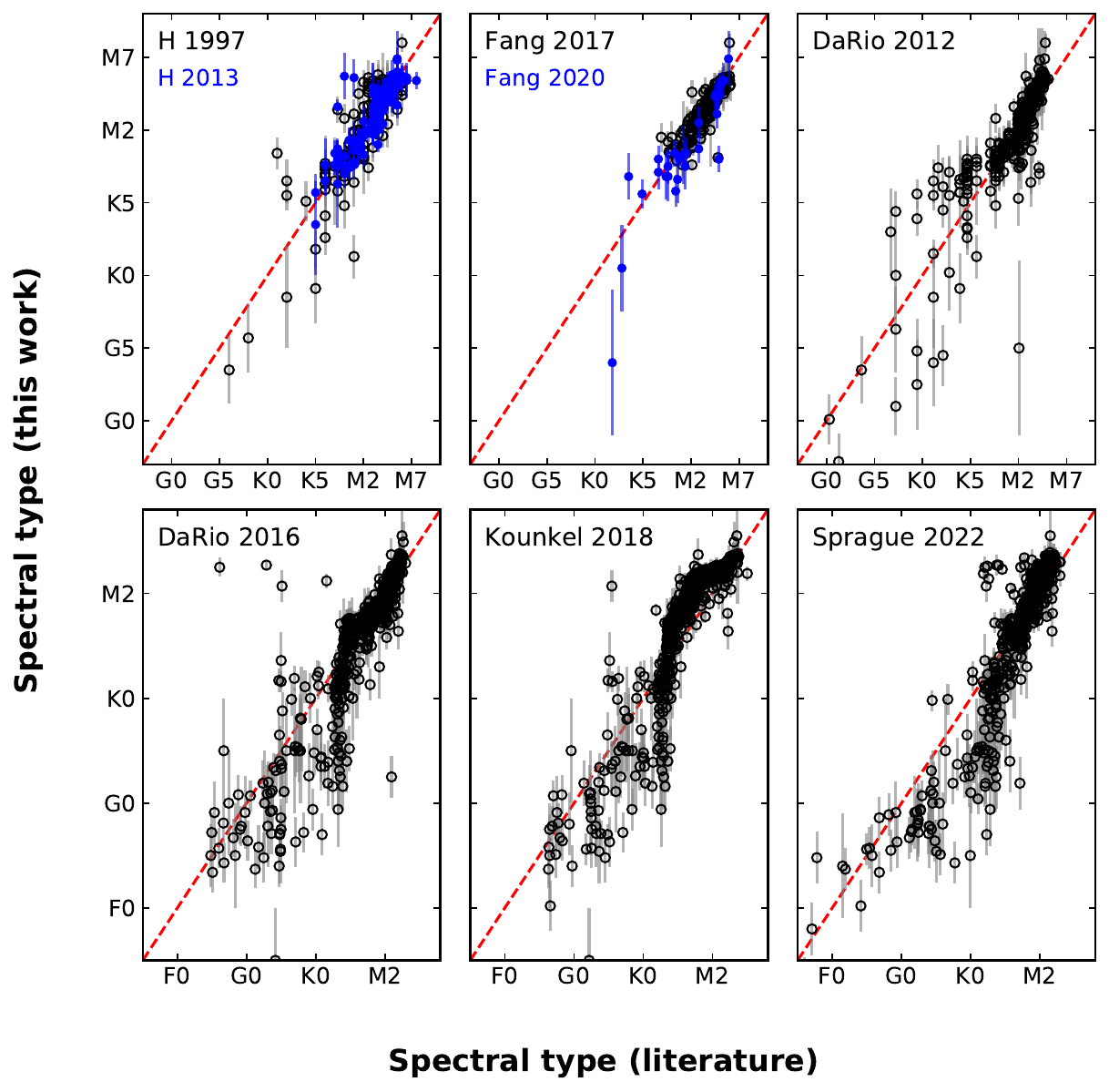}
    \caption{Comparison between the spectral types derived in this work and those reported in previous studies: H 1997 and H 2013 \citep{Hillenbrand1997,Hillenbrand2013}, DaRio 2012 \citep{DaRio2012}, Fang 2017 and Fang 2021 \citep{Fang2017,Fang2021}, DaRio 2016 \citep{DaRio2016}, Kounkel 2018 \citep{Kounkel2018} and Sprague 2022 \citep{Sprague2022}. Vertical error bars represent the uncertainties associated with our spectral type classification. The solid line indicates the one-to-one relation. For most stars, the spectral types derived in this work agree with those reported in the literature within the uncertainties.}
    \label{fig:SpT_comp}
\end{figure*}

We compared our \sptclass spectral-type estimates with those reported in previous studies that derived \teff or spectral types using different datasets and methodologies (Figure~\ref{fig:SpT_comp}).\\

An optical low-resolution (R $\sim$ 1500-2000) spectroscopic census of the ONC was presented by \citet{Hillenbrand2013}, updating the earlier work of \citet{Hillenbrand1997} and providing spectral types or spectral-type constraints for 1,526 photometric ONC candidates. They employed spectroscopic indices based on TiO and VO molecular bands for K- and M-type stars, and atomic-line indices for spectral types earlier than K0.\\

\citet{Fang2017} presented a spectroscopic census based primarily on Hectospec observations, deriving extinctions and stellar masses for 691 ONC members. Using a methodology similar to that implemented in \sptclass and calibrated with X-Shooter spectra of a sample of known YSOs, they constructed empirical spectroscopic relations to estimate spectral types. 
In a subsequent study, \citet{Fang2021} used integral-field spectroscopic data obtained with the Multi Unit Spectroscopic Explorer (MUSE) at the Very Large Telescope to derive spectral types by fitting observed spectra with templates constructed from X-Shooter data over the wavelength range $\sim$5500 - 9000 \AA. They employed a Markov Chain Monte Carlo (MCMC) approach to simultaneously estimate spectral types, visual extinctions, and accretion-related veiling by maximising the likelihood function.\\

\citet{DaRio2012} estimated \teff using a reddening-independent photometric index combining wide I-band photometry with two medium-band filters centered at 753 and 770~nm, which measure the depth of TiO molecular absorption features in low-mass stars. They derived \teff, visual extinctions, luminosities, and stellar masses for 1,743 stars within a $\sim$900 arcmin$^2$ region of the ONC.\\

Using high-resolution (R $\sim$ 22,500) H-band spectroscopy from the APOGEE survey of the Sloan Digital Sky Survey, \citet{DaRio2016}, \citet{Kounkel2018}, and \citet{Sprague2022} estimated \teff for large samples of ONC stars using different analysis techniques. \citet{DaRio2016} employed a pipeline that compares APOGEE spectra with the BT-Settl synthetic spectral grid, deriving best-fit estimates of radial velocity (RV), projected rotational velocity (\vsini), \logg, and \teff using an MCMC approach. \citet{Kounkel2018} presented a spectroscopic census of the Orion star-forming complex, including the ONC, and estimated RV, \vsini, \logg, \teff, and veiling by fitting APOGEE spectra with PHOENIX model atmospheres, also using an MCMC approach. 

Finally, \citet{Olney2020} developed the APOGEEnet tool, a convolutional neural network (CNN) trained to predict stellar parameters (e.g., \teff, \logg, and [Fe/H]) from APOGEE spectra. \citet{Sprague2022} subsequently extended the APOGEEnet tool to model a wide range of stellar types, including low-mass main-sequence stars, red giants, PMS stars, and OB-type stars. The version presented by \citet{Sprague2022} was trained using a heterogeneous set of labels derived from photometric relations, theoretical isochrones, literature stellar parameters, and spectral-type/luminosity-class calibrations for high-mass stars, enabling the determination of stellar parameters across a much broader region of parameter space than the original APOGEEnet implementation.

Figure~\ref{fig:SpT_comp} compares our spectral-type estimates with those reported in the previous studies. In general, the literature classifications are consistent, within the uncertainties, with those derived using \sptclass in this work. The closest agreement is found with the spectral types reported by \citet{Fang2017}, who defined indices of individual spectral features in a manner similar to the \sptclass methodology. 
We find similarly close agreement with the classifications reported by \citet{Hillenbrand2013} for late type stars, 
but the largest discrepancies are found for three early-K stars, which we classify more than four subtypes later. Based on the strength of molecular absorption bands such as TiO and CaH, we confirm that these stars have spectral types of K5 or later.

We converted the \teff derived from APOGEE spectra (lower panels of Figure~\ref{fig:SpT_comp}) into spectral types using the empirical relations of \citet{Pecaut2013}. Although our spectral types show a clear correlation with those from the APOGEE studies, and most stars agree within their uncertainties,
we observe substantial discrepancies particularly among early-K stars. In particular, we note a systematic tendency for the temperatures derived from APOGEE spectra \citep{DaRio2016, Kounkel2018, Sprague2022} to yield later spectral types than those derived from our optical spectra, especially for early-K stars. This behaviour is consistent with recent comparisons between optical and near-infrared spectral classifications of PMS stars, which suggest that cool stellar spots can introduce systematic differences between temperatures derived from optical and infrared spectra \citep[e.g.,][]{Flores2022,Gangi2022,Perez2024}. 
Because the contrast between spotted and unspotted photospheric regions is larger at optical wavelengths than in the near-infrared, \teff derived from optical spectra tend to be systematically hotter than those inferred from infrared spectra. This effect is expected to be most pronounced for stars with temperatures of approximately 4500--5000 K (early-K spectral types), where the largest offsets between optical and infrared spectral classifications have been reported. Part of the observed scatter may also arise from differences in the methodologies used to derive stellar parameters, including the conversion between \teff and spectral type and the different spectral diagnostics employed at optical and near-infrared wavelengths. 
Together, these effects may contribute to the largest discrepancies observed between our classifications and those derived from APOGEE spectra.

\subsection{Kinematic candidates}
\label{sec:kinmem}

\begin{figure}
    \centering
    \includegraphics[width=0.8\columnwidth]{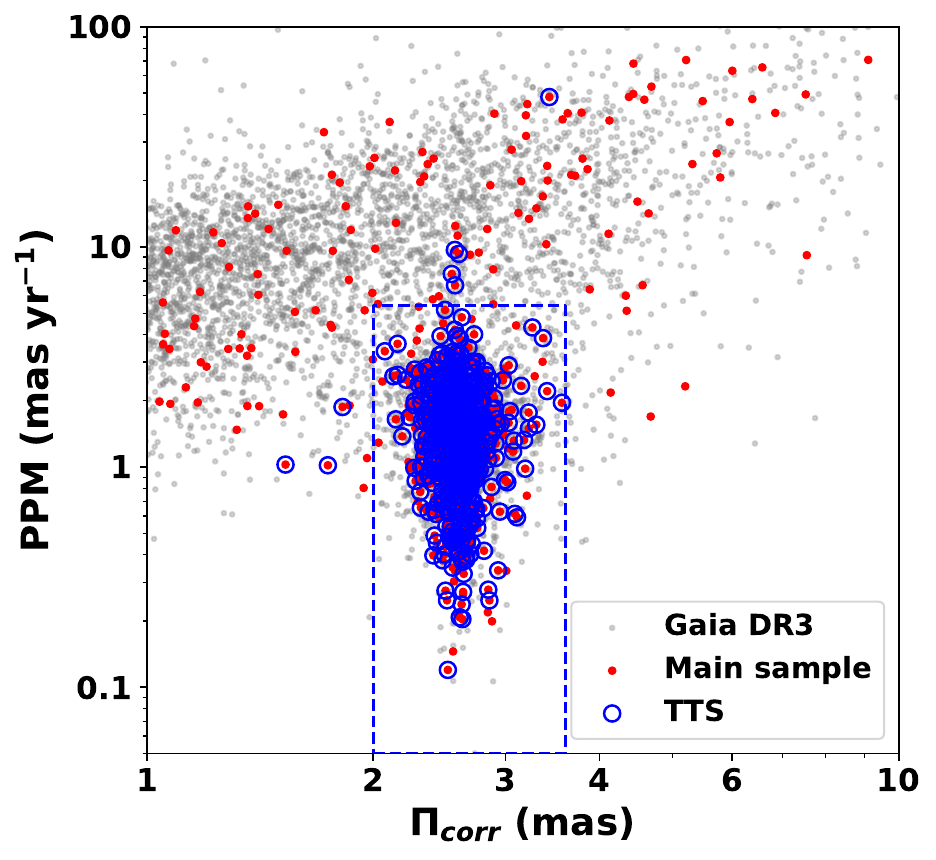}
    \caption{
    Parallaxes versus proper-motion modulus for \gaia sources that satisfy the astrometric quality criterion described in Section~\ref{sec:kinmem} (grey dots). Red circles represent stars from the main sample analysed in this work, while blue circles indicate spectroscopically identified TTS (spec\_flag = TTS, CTTS, WTTS, or CWTTS). The blue lines mark the kinematic selection region}
    \label{fig:KINMEM}
\end{figure}

\gaia provides high-precision astrometric data that enable robust kinematic membership analyses of stellar clusters formed in recent star-forming events \citep[e.g.,][]{Hunt2023,Perren2023}. We applied a methodology similar to that described by \citet{Sergio2024} to identify PMS kinematic candidates in the Orion star-forming complex. Briefly, using parallaxes corrected for systematics \citep[$\Pi_{corr}$; ][]{Lindegren2021b}\footnote{$\Pi_{corr}$ is obtained using the zero-point bias obtained from the ARI Gaia
TAP service}, we applied three selection criteria to \gaia sources in the ONC region:
\begin{enumerate}
\item Quality cut: We selected stars with relative parallax uncertainties better than 20\% and satisfying at least one of two commonly adopted astrometric quality criteria, namely, a renormalized unit weight error (RUWE) smaller than 1.4 or a fidelity\_v2 parameter greater than 0.5. The RUWE evaluates the quality of the astrometric solution under the assumption of single-star motion \citep{Lindegren2021a}, while the fidelity\_v2 parameter provides a machine-learning-based assessment of astrometric reliability \citep{Rybizki2022}.

\item PMS cut: We selected 
PMS candidates younger than 30 Myr using the criterion G$_\mathrm{RP}$-5 $\log(1000/\Pi_{corr})$+5 $<$ 2.6 + 2.1 (G$_\mathrm{BP}$-G$_\mathrm{RP}$), which corresponds to a linear approximation of the 30 Myr isochrone derived from the PARSEC evolutionary models \citep{Marigo2017}. The reddening vector in this diagram is nearly parallel to the adopted linear boundary, making this criterion relatively insensitive to extinction and preventing the exclusion of YSO candidates with low or moderate extinction \citep{Sergio2024}.

\item Kinematic cut: We selected stars with proper-motion modulus (PMM) $<$ 5.425 mas \yrinv and $\Pi_{corr}$ between 2 and 3.6 mas (Figure~\ref{fig:KINMEM}). These thresholds were derived from the distribution of spectroscopically confirmed TTS in the Orion star-forming complex \citep{Briceno2019, Hernandez2023} using the median and the median absolute deviation (MAD = $\sigma$/1.4826). The adopted limits correspond to a 3$\sigma$ interval in parallax and a 6$\sigma$ upper limit in PMM, assuming Gaussian statistics (1 $\sigma \approx 1.4826\times$MAD). The relatively permissive PMM threshold was adopted to avoid rejecting YSOs with atypical kinematics produced by dynamical interactions during the star-formation process (e.g., runaway or walkaway TTS).

\end{enumerate}

Stars satisfying all three criteria were flagged as kinematic candidates (k\_flag = k\_mem) in Table~\ref{tab:main}. Stars that meet the quality criterion but fail either the PMS or kinematic cuts were classified as kinematic non-members (k\_flag = k\_nmem). Objects lacking \gaia parallaxes or failing the quality criterion were labelled as having uncertain membership (k\_flag = k\_unc). \\

Out of 1,513 stars in the main sample, 
1,069 were classified as kinematic candidates. In addition, we incorporated the kinematic grouping information reported by \citet{Sergio2024}, who identified kinematic structures using a clustering algorithm. Among our kinematic members, 712 stars were assigned to the big structures identified in that study, with 707 belonging to the ONC and 5 associated with the OBP-Far structure, a population located approximately 33 pc farther away.\\

To define a bona fide sample of TTS, we combined the kinematic selection with the spectroscopic classification described in Section~\ref{sec:sptclass} and selected kinematic candidates with spec\_flag = TTS, CTTS, WTTS, or CWTTS, resulting in a sample of 780 bona fide TTS.\\

Applying the three criteria described above, we identified a total of 2,804 kinematic candidates in the ONC region. Of these, 1,069 belong to the main sample, while the remaining 1,735 do not have optical spectra analysed in this work. As a complementary catalogue, we compiled spectral types and \teff available in the literature for these 1,735 sources and summarized the results in Appendix~\ref{ap:kinmem_teff}. When only \teff were reported \citep[e.g.,][]{Olney2020,DaRio2012, DaRio2016}, we converted them into spectral types using the calibration of \citet{Pecaut2013}.

\section{\massage code: reddening, masses, and ages}\label{section:Massage}

The \massage code is a Python-based tool{\footnote{Initially an IDL-based code}} that estimates extinction, luminosity, mass, \logg, and age for a sample of stars by comparing their observed positions in the HRD with predictions from stellar evolutionary models \citep[e.g.,][]{Serna2021,Bonilla2022,Bonilla2024,Ricardo2023,Ricardo2025,Carlos2023,Hernandez2023}. 
To estimate these parameters, the code requires the following inputs: spectral types or \teff, parallaxes or distances, \gaiam photometry (G$_\mathrm{BP}$, G, and G$_\mathrm{RP}$), and the J and H magnitudes from the 2MASS catalogue. \massage also requires uncertainties in the inputs, which are propagated to the derived output quantities using a Monte Carlo (MC) error-propagation approach \citep{Anderson1976}, assuming Gaussian distributions for the input uncertainties. 
\massage follows three main steps to derive visual extinctions, bolometric luminosities, and stellar masses and ages. These steps, together with the available configuration options, are described in Appendix~\ref{ap:massage_procedure}, which also discusses the main caveats and limitations of the code. The Python implementation of \massage will be made publicly available upon publication at \url{https://github.com/jesush149/MassAge_py}.
Below, we describe the options adopted for the main sample.

First, we obtained bolometric corrections and converted spectral types to \teff using the calibrations from \citet{Pecaut2013}. We used the standard colours provided by \citet{Luhman2020} and adopted the extinction law of \citet{CCM1989} with the standard interstellar value for the total-to-selective extinction ratio ($R_V$ = 3.1). Although relatively high values of $R_V$ have been reported for the ONC \citep[$R_V$ = 5.5 $\pm$ 0.7;][]{Costero1970}, we adopted the standard value, which better reproduces the observed broadband colours of ONC members \citep{DaRio2012}. 

We selected the J band as the anchor filter because it is less affected by extinction than optical bands while remaining largely photospheric for most sources and being less affected by infrared excess emission than longer-wavelength bands. Accordingly, the standard and observed colours are defined as $M_\lambda$-J, where $M_\lambda$ denotes the remaining magnitudes used by \massage (Equation~\ref{e:extinction}). 

To account for the strong and highly variable reddening in the ONC, where visual extinctions range from nearly negligible values up to \av$\gtrsim$6 mag \citep[e.g.,][]{Scandariato2011}, we selected the option to use effective filter wavelengths rather than the nominal filter wavelengths. 
In this approach, the effective wavelength is obtained by 
integrating the expected stellar spectrum weighted by
the filter transmission curve \citep[Equation~\ref{e:WLeff};][]{Bessell2012}. 
A similar approach was adopted by \citet{Alzate2023} to investigate the star-formation history of the ONC using \gaia, although those authors employed the extinction-coefficient polynomials provided by the \gaiam Collaboration, which account for both the source spectral energy distribution (SED) and the amount of extinction.

\begin{figure}
    \centering
    \includegraphics[width=0.9\columnwidth]{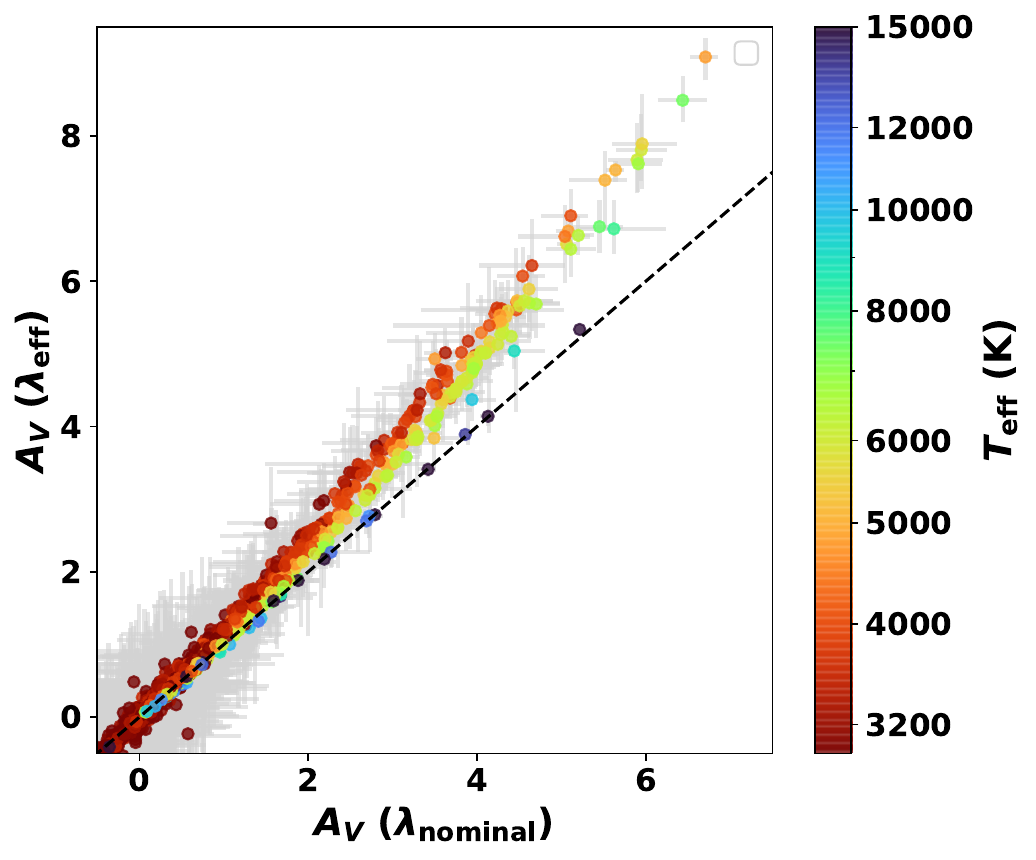}
    \caption{Comparison between extinction estimates derived using nominal filter wavelengths and those obtained using effective wavelengths that account for the intrinsic SED of the source and the filter transmission curves. The one-to-one relation is shown for reference (dashed line). The colour bar indicates the effective temperature of the sources.}
    \label{fig:av_comp}
\end{figure}

 Figure~\ref{fig:av_comp} compares the extinction estimated for the main sample using the nominal wavelengths of the photometric filters with that obtained using the effective filter wavelengths.
 In general, Figure~\ref{fig:av_comp} shows that the use of nominal filter wavelengths systematically underestimates \av.
 The differences are systematically larger for stars with larger \av, highlighting the importance of accounting for wavelength-dependent reddening when estimating extinction in embedded regions such as the ONC using broad-band filters. Table~\ref{tab:main} includes the \teff derived using the calibrations of \citet{Pecaut2013}, the \av values obtained using the effective-wavelength approach, and the stellar luminosities derived with the \massage code.\\


\begin{figure*}
    \centering
    \includegraphics[width=0.9\textwidth]{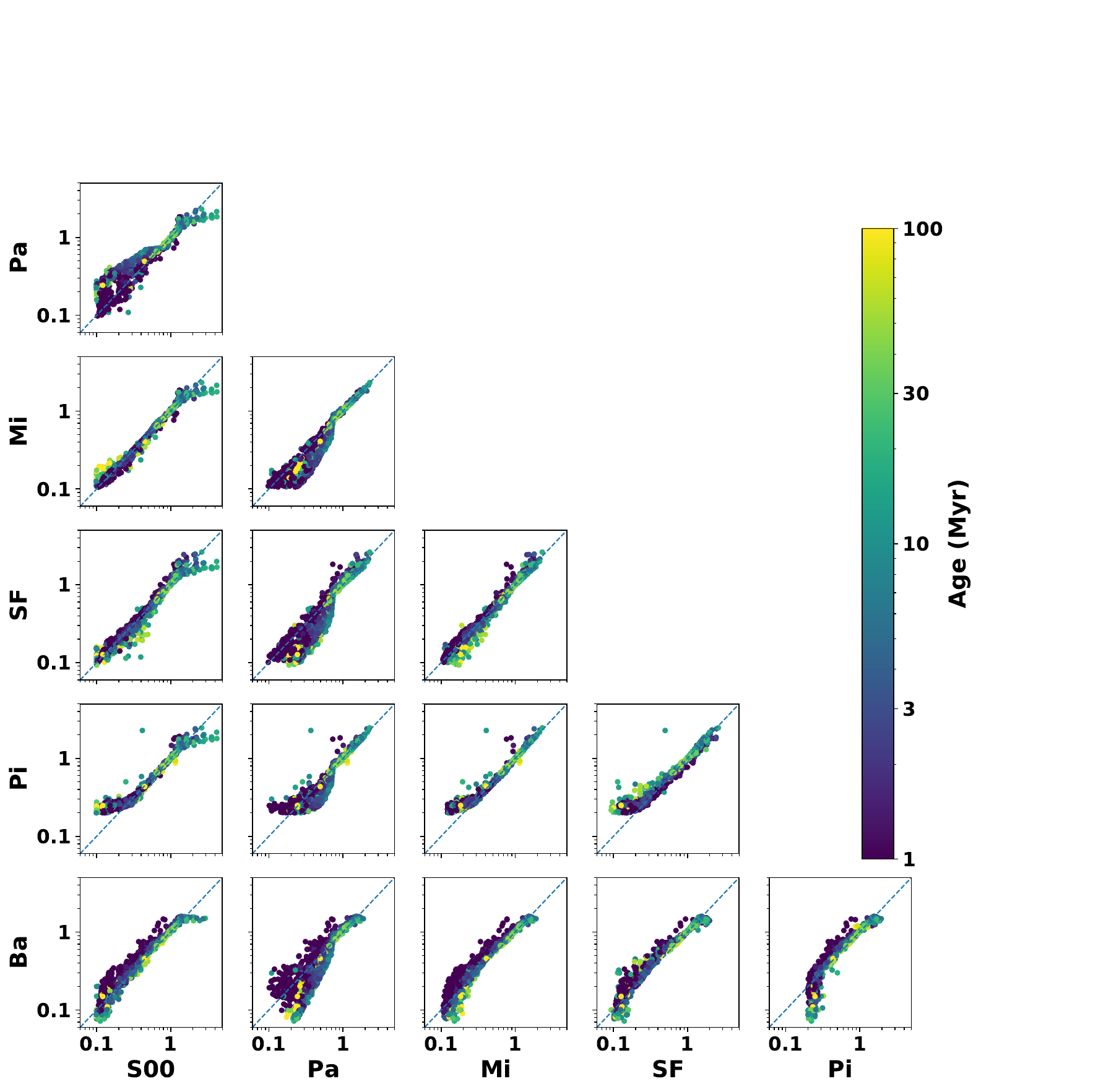}
    \caption{Comparison of stellar masses derived with the \massage code using different non-magnetic evolutionary models. Each panel compares a pair of models, with the one-to-one relation shown for reference. Stellar masses are expressed in units of solar masses (\msun). The abbreviations denote the evolutionary models adopted: Pa = PARSEC \citep{Marigo2017}, Mi = MIST \citep{Dotter2016}, SF = \citet{Siess2000}, Pi = Pisa \citep{Tognelli2011}, Ba = \citet{Baraffe2002,Baraffe2015}, and S00 = \citet{Somers2020} without stellar spots. The colour of each point indicates the stellar ages from S00, as shown by the colour bar.}
    \label{fig:mass1}
\end{figure*}

\begin{figure*}
    \centering
    \includegraphics[width=0.9\textwidth]{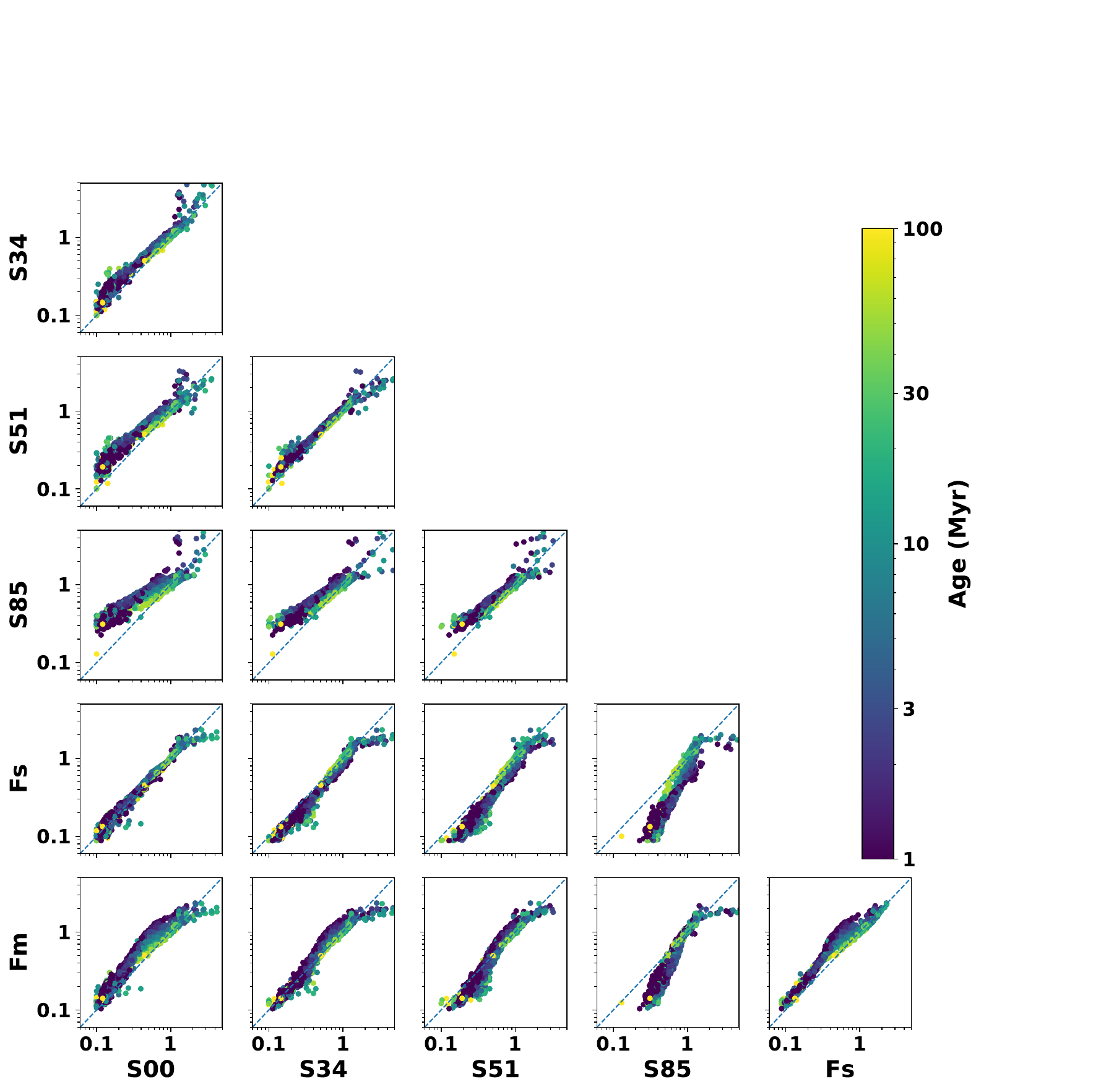}
    \caption{Comparison of stellar masses derived with the \massage code using evolutionary models that incorporate magnetic fields or stellar spots. Each panel compares a pair of models, with the one-to-one relation shown for reference. Stellar masses are expressed in units of solar masses (\msun). The abbreviations denote the evolutionary models adopted: Fs = \citet{Feiden2016} non-magnetic model, Fm = \citet{Feiden2016} magnetic model, S00 = \citet{Somers2020} without stellar spots, and S34, S51, and S84 = \citet{Somers2020} models with 34\%, 51\%, and 85\% spot coverage, respectively.The colour of each point indicates the stellar ages from S00, as shown by the colour bar.}
    \label{fig:mass2}
\end{figure*}

\begin{figure*}
    \centering
    \includegraphics[width=0.9\textwidth]{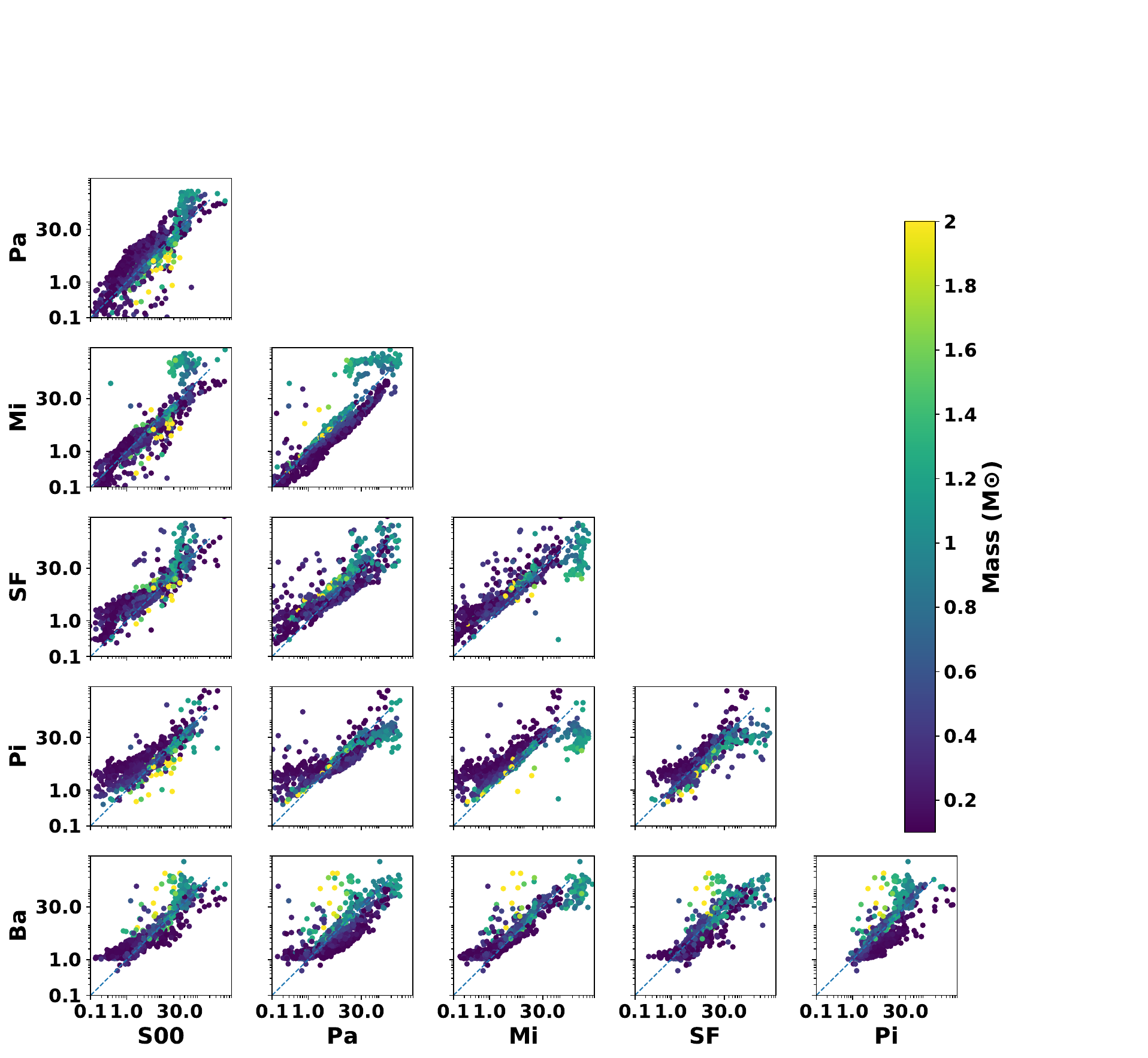}
    \caption{
    Comparison of stellar ages derived with the \massage code using different non-magnetic evolutionary models. Each panel compares a pair of models, with the one-to-one relation shown for reference. Stellar ages are expressed in Myr. The abbreviations denote the evolutionary models adopted: Pa = PARSEC \citep{Marigo2017}, Mi = MIST \citep{Dotter2016}, SF = \citet{Siess2000}, Pi = Pisa \citep{Tognelli2011}, Ba = \citet{Baraffe2002,Baraffe2015}, and S00 = \citet{Somers2020} without stellar spots. The colour of each point indicates the stellar masses from S00, as shown by the colour bar.}
    \label{fig:age1}
\end{figure*}

\begin{figure*}
    \centering
    \includegraphics[width=0.9\textwidth]{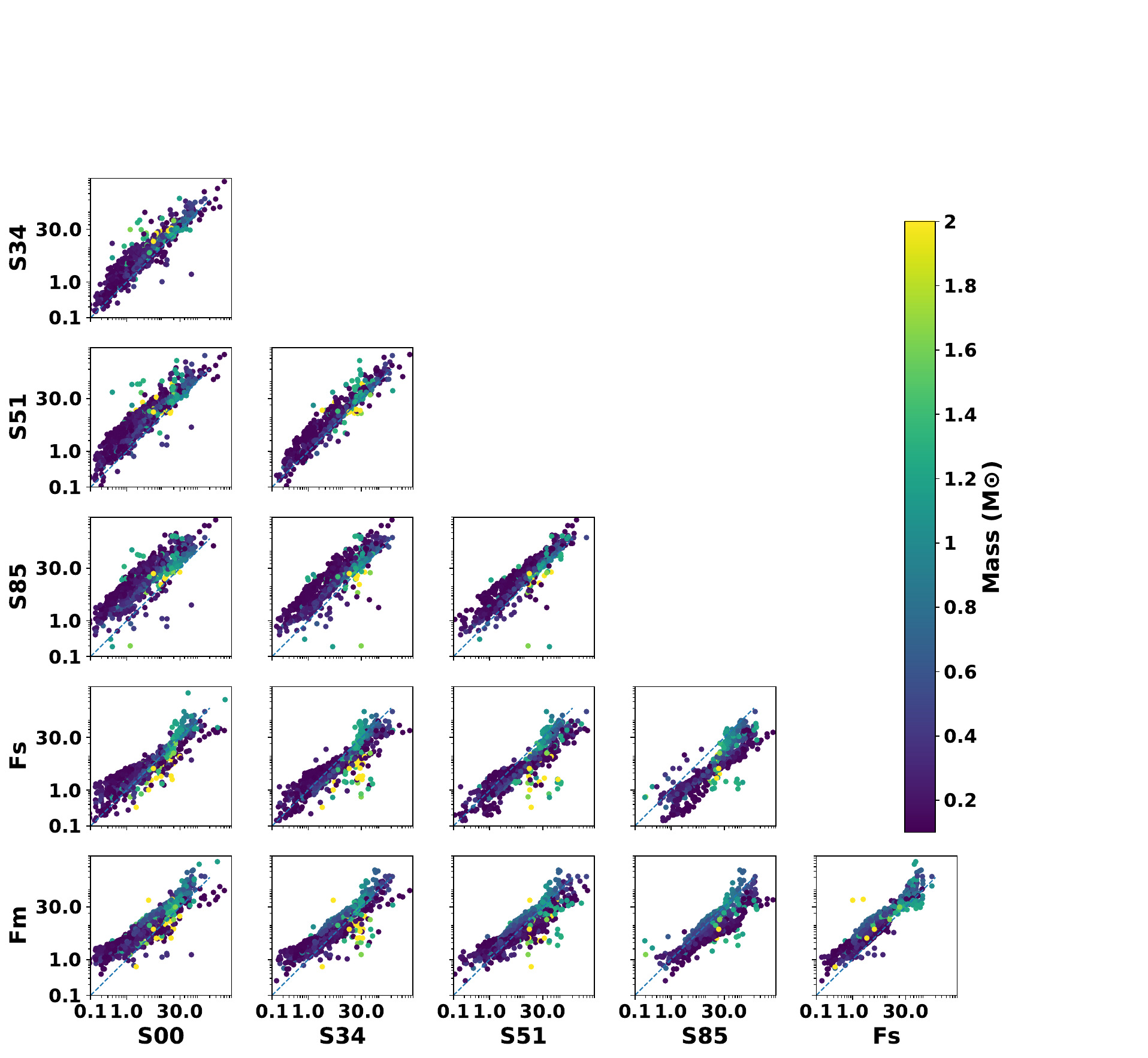}
    \caption{Comparison of stellar ages derived with the \massage code using evolutionary models that incorporate magnetic fields or stellar spots. Each panel compares a pair of models, with the one-to-one relation shown for reference. Stellar ages are expressed in Myr. The abbreviations denote the evolutionary models adopted: Fs = \citet{Feiden2016} non-magnetic model, Fm = \citet{Feiden2016} magnetic model, S00 = \citet{Somers2020} without stellar spots, and S34, S51, and S84 = \citet{Somers2020} models with 34\%, 51\%, and 85\% spot coverage, respectively. The colour of each point indicates the stellar masses from S00, as shown by the colour bar.}
    \label{fig:age2}
\end{figure*}

\begin{table}
\centering
\caption{Masses and ages (non-magnetic models) }
\begin{tabular}{ll}
\hline
Column label & Description       \\
\hline \hline
Gaia DR3\_ID   &  \gaia source name \\
mass\_Pa & median mass from \citet{Marigo2017}\\
mass\_Pa low & lower limit of mass from \citet{Marigo2017}\\
mass\_Pa upp & upper limit of mass from \citet{Marigo2017} \\
age\_Pa & median age from \citet{Marigo2017}\\
age\_Pa low & lower limit of age from \citet{Marigo2017}\\
age\_Pa upp & upper limit of age from \citet{Marigo2017} \\
flag\_Pa & \massage's flag from \citet{Marigo2017} \\
mass\_Mi & median mass from \citet{Dotter2016}\\
mass\_Mi low & lower limit of mass from \citet{Dotter2016}\\
mass\_Mi upp & upper limit of mass from \citet{Dotter2016} \\
age\_Mi & median age from \citet{Dotter2016}\\
age\_Mi low & lower limit of age from \citet{Dotter2016}\\
age\_Mi upp & upper limit of age from \citet{Dotter2016} \\
flag\_Mi & MassAge's flag from \citet{Dotter2016} \\
mass\_SF & median mass from \citet{Siess2000}\\
mass\_SF low & lower limit of mass from \citet{Siess2000}\\
mass\_SF upp & upper limit of mass from \citet{Siess2000} \\
age\_SF & median age from \citet{Siess2000}\\
age\_SF low & lower limit of age from \citet{Siess2000}\\
age\_SF upp & upper limit of age from \citet{Siess2000} \\
flag\_SF & \massage's flag from \citet{Siess2000} \\
mass\_Pi & median mass from \citet{Tognelli2011}\\
mass\_Pi low & lower limit of mass from \citet{Tognelli2011}\\
mass\_Pi upp & upper limit of mass from \citet{Tognelli2011} \\
age\_Pi & median age from \citet{Tognelli2011}\\
age\_Pi low & lower limit of age from \citet{Tognelli2011}\\
age\_Pi upp & upper limit of age from \citet{Tognelli2011} \\
flag\_Pi & \massage's flag from \citet{Tognelli2011} \\
mass\_Ba & median mass from \citet{Baraffe2002,Baraffe2015}\\
mass\_Ba low & lower limit of mass from \citet{Baraffe2002,Baraffe2015}\\
mass\_Ba upp & upper limit of mass from \citet{Baraffe2002,Baraffe2015} \\
age\_Ba & median age from \citet{Baraffe2002,Baraffe2015}\\
age\_Ba low & lower limit of age from \citet{Baraffe2002,Baraffe2015}\\
age\_Ba upp & upper limit of age from \citet{Baraffe2002,Baraffe2015} \\
flag\_Ba & \massage's flag from \citet{Baraffe2002,Baraffe2015} \\
mass\_Fs & median mass from \citet{Feiden2016}, non-magnetic\\
mass\_Fs low & lower limit of mass from \citet{Feiden2016}, non-magnetic\\
mass\_Fs upp & upper limit of mass from \citet{Feiden2016}, non-magnetic \\
age\_Fs & median age from \citet{Feiden2016}, non-magnetic \\
age\_Fs low & lower limit of age from \citet{Feiden2016}, non-magnetic \\
age\_Fs upp & upper limit of age from \citet{Feiden2016}, non-magnetic \\
flag\_Fs & \massage's flag from \citet{Feiden2016}, non-magnetic \\
mass\_S00 & median mass from \citet{Somers2020}, unspotted\\
mass\_S00 low & lower limit of mass from \citet{Somers2020}, unspotted \\
mass\_S00 upp & upper limit of mass from \citet{Somers2020}, unspotted \\
age\_S00 & median age from  \citet{Somers2020}, unspotted\\
age\_S00 low & lower limit of age from \citet{Somers2020}, unspotted \\
age\_S00 upp & upper limit of age from \citet{Somers2020}, unspotted \\
flag\_S00 & \massage's flag from \citet{Somers2020}, unspotted \\
\\
\hline
\end{tabular}
\begin{tablenotes}
\item Lower and upper limits represent the 16th and 84th percentiles, enclosing $\sim$68\% of the distribution for a Gaussian case.
\end{tablenotes}
\label{tab:mass_and_ages}
\end{table}

\begin{table}
\centering
\caption{Masses and ages (magnetic models) }
\begin{tabular}{ll}
\hline
Column label & Description       \\
\hline \hline
Gaia DR3\_ID   &  \gaia source name \\
mass\_Fm & median mass from \citet{Feiden2016}, magnetic\\
mass\_Fm low & lower limit of mass from \citet{Feiden2016}, magnetic\\
mass\_Fm upp & upper limit of mass from \citet{Feiden2016}, magnetic\\
age\_Fm & median age from \citet{Feiden2016}, magnetic\\
age\_Fm low & lower limit of age from \citet{Feiden2016}, magnetic\\
age\_Fm upp & upper limit of age from \citet{Feiden2016}, magnetic \\
flag\_Fm upp & \massage's flag from \citet{Feiden2016}, magnetic \\
mass\_S17 & median mass \citep[][ 17\% spot ]{Somers2020}\\
mass\_S17 low & lower limit of mass \citep[][ 17\% spot ]{Somers2020} \\
mass\_S17 upp & upper limit of mass \citep[][ 17\% spot ]{Somers2020} \\
age\_S17 & median age \citep[][ 17\% spot ]{Somers2020}\\
age\_S17 low & lower limit of age\citep[][ 17\% spot ]{Somers2020}\\
age\_S17 upp & upper limit of age \citep[][ 17\% spot ]{Somers2020}\\
flag\_S17 upp & \massage's flag \citep[][ 17\% spot ]{Somers2020} \\
mass\_S34 & median mass \citep[][ 34\% spot ]{Somers2020}\\
mass\_S34 low & lower limit of mass \citep[][ 34\% spot ]{Somers2020}\\
mass\_S34 upp & upper limit of mass \citep[][ 34\% spot ]{Somers2020}\\
age\_S34 & median age \citep[][ 34\% spot ]{Somers2020}\\
age\_S34 low & lower limit of age \citep[][ 34\% spot ]{Somers2020}\\
age\_S34 upp & upper limit of age \citep[][ 34\% spot ]{Somers2020}\\
flag\_S34 upp & \massage's flag \citep[][ 34\% spot ]{Somers2020} \\
mass\_S51 & median mass \citep[][ 51\% spot ]{Somers2020}\\
mass\_S51 low & lower limit of mass \citep[][ 51\% spot ]{Somers2020}\\
mass\_S51 upp & upper limit of mass \citep[][ 51\% spot ]{Somers2020}\\
age\_S51 & median age \citep[][ 51\% spot ]{Somers2020}\\
age\_S51 low & lower limit of age \citep[][ 51\% spot ]{Somers2020}\\
age\_S51 upp & upper limit of age \citep[][ 51\% spot ]{Somers2020}\\
flag\_S51 upp & \massage's flag \citep[][ 51\% spot ]{Somers2020} \\
mass\_S68 & median mass \citep[][ 68\% spot ]{Somers2020}\\
mass\_S68 low & lower limit of mass \citep[][ 68\% spot ]{Somers2020}\\
mass\_S68 upp & upper limit of mass \citep[][ 68\% spot ]{Somers2020}\\
age\_S68 & median age \citep[][ 68\% spot ]{Somers2020}\\
age\_S68 low & lower limit of age \citep[][ 68\% spot ]{Somers2020}\\
age\_S68 upp & upper limit of age \citep[][ 68\% spot ]{Somers2020}\\
flag\_S68 upp & \massage's flag \citep[][ 68\% spot ]{Somers2020} \\
mass\_S85 & median mass \citep[][ 85\% spot ]{Somers2020}\\
mass\_S85 low & lower limit of mass \citep[][ 85\% spot ]{Somers2020}\\
mass\_S85 upp & upper limit of mass \citep[][ 85\% spot ]{Somers2020}\\
age\_S85 & median age \citep[][ 85\% spot ]{Somers2020}\\
age\_S85 low & lower limit of age \citep[][ 85\% spot ]{Somers2020}\\
age\_S85 upp & upper limit of age \citep[][ 85\% spot ]{Somers2020}\\
flag\_S85 upp & \massage's flag \citep[][ 85\% spot ]{Somers2020} \\
\\
\hline
\end{tabular}
\begin{tablenotes}
\item Lower and upper limits represent the 16th and 84th percentiles, enclosing $\sim$68\% of the distribution for a Gaussian case.
\end{tablenotes}
\label{tab:mass_and_ages2}
\end{table}

Finally, we estimated stellar masses and ages for the main sample using different evolutionary models, adopting cubic interpolation on the HRD and performing 300 MC iterations, which provide stable estimates of the output uncertainties while maintaining a reasonable computational cost. Figures~\ref{fig:mass1}, \ref{fig:mass2}, \ref{fig:age1} and~\ref{fig:age2} compare the masses and ages derived from the different theoretical grids. Tables~\ref{tab:mass_and_ages} and~\ref{tab:mass_and_ages2} present the stellar masses and ages derived using the \massage code for the main sample based on different evolutionary models.

In general, different evolutionary models yield systematically different masses and ages. For example, magnetic stellar evolution models from \citet{Feiden2016} predict systematically older ages and higher stellar masses than their non-magnetic counterparts. Similarly, models that include significant star-spot coverage, such as those presented by \citet{Somers2020}, tend to yield systematically older ages and higher masses, owing to the impact of spots on stellar structure and emergent luminosity. 
These results are consistent with those reported by \citet{Alzate2023}, who found a constant star-formation rate between 0.3 and 5 Myr when using PARSEC and MIST models, while the inferred age range extends to approximately 10 Myr when using models that include stellar spots \citep{Somers2020}. The systematic differences between spotted and unspotted evolutionary models should be interpreted in conjunction with the observational effects discussed in Section~\ref{sec:sptclass}, where stellar spots were also identified as a potential source of systematic uncertainty in spectral classifications.

Even among standard (non-magnetic, unspotted) evolutionary models, notable systematic differences are present (Figures~\ref{fig:mass1} and~\ref{fig:age1}). In particular, the MIST stellar evolution models \citep{Dotter2016} tend to produce systematically older ages than the PARSEC models \citep{Marigo2017}, especially for young ($<$30 Myr) low-mass stars. The stellar masses derived from these two model grids show generally good agreement for stars more massive than $\sim$0.75 \msun. However, below this limit, the PARSEC models tend to yield systematically higher stellar masses than the MIST models.
These differences highlight the model dependence of stellar mass and age estimates derived from HRD, particularly for young low-mass stars. While determining which evolutionary model provides a more accurate description of PMS evolution is beyond the scope of this work, it is important to adopt a consistent model when performing statistical comparisons between different studies or stellar populations.

\section{Infrared Analysis}
\label{sec:infrared}

We carried out a census of protoplanetary discs for the main sample using infrared photometry from the IRAC and MIPS instruments aboard the \spitzer Space Telescope \citep{Fazio2004, Rieke2004}. The observational setup and aperture photometry procedures are described by \citet{Megeath2012}. In this work, we use the full \spitzer source catalogue associated with that survey, including sources that were excluded from the YSOs catalogue of \citet{Megeath2012} by their selection criteria (hereafter the full IR detection sample).

\subsection{Infrared Excess Identification}

\begin{figure}
    \centering
    \includegraphics[width=0.9\columnwidth]{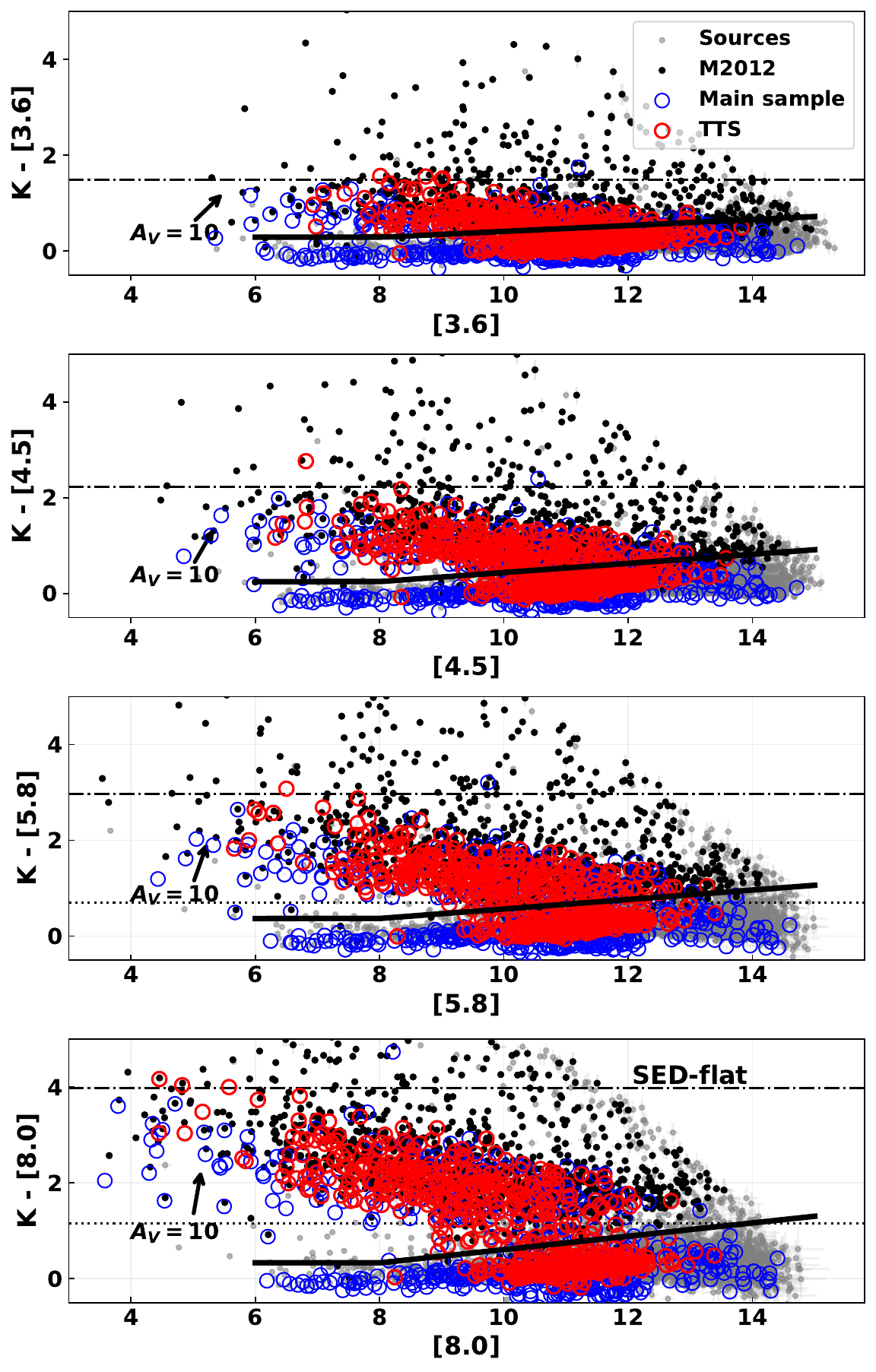}
    \caption{Colour-magnitude diagrams 
    used to identify stars with infrared excesses at 5.8 and 8.0~\micron. YSOs from \citet{Megeath2016} are shown as black dots. Stars in the main sample are shown as blue open circles, and bona fide T Tauri stars, which constitute a subset of the main sample, are highlighted with red open circles.    
    The photospheric limits (black solid lines) were derived from Gaussian fits to the K-[3.6], K-[4.5], K-[5.8], and K-[8.0] colour distributions. 
    The lower limits for CII objects from \citet{Luhman2010} (dotted lines) and the CII/CI boundaries corresponding to a flat SED (dashed-dotted lines) are also shown.}
    \label{fig:IRACclass}
\end{figure}

Following the classical infrared classification scheme of young stellar objects \citep{Lada1987}, we adopt colour-based proxies to distinguish sources with little or no infrared excess (Class III; hereafter CIII), stars surrounded by optically thick circumstellar discs (Class II; hereafter CII), and embedded protostars with substantial envelope emission (Class I; hereafter CI). Although these categories were originally defined using the infrared spectral index measured between 2 and 20 \micron, the colour criteria adopted here are intended to provide approximate equivalents to the infrared classes.

Figure~\ref{fig:IRACclass} shows the colour–magnitude diagrams used to identify stars with infrared excesses at 3.6, 4.5, 5.8, and 8.0~\micron. YSOs from \citet{Megeath2012} are shown as black dots. The infrared excesses for the main sample were independently identified in each of the IRAC bands and at 24~\micron\ using reddening-corrected colours. The photospheric limits (solid lines) were derived by dividing the full IR detection sample into 0.5 mag bins in the [3.6], [4.5], [5.8], and [8.0] IRAC bands and fitting Gaussian functions to the colour distributions (K–[3.6], K–[4.5], K–[5.8], and K–[8.0]) after applying a 3$\sigma$ clipping. We then estimated the Gaussian centre and standard deviation of the colour distributions in each magnitude bin.

In the faint regime (IRAC magnitudes $[\mathrm{IRAC}]\geq 8$~mag), we fitted straight lines to the upper envelopes defined by the Gaussian centres plus $3\sigma$ in the K-[3.6], K-[4.5], K-[5.8], and K-[8.0] colour–magnitude diagrams (Figure~\ref{fig:IRACclass}). The Bayesian Information Criterion (BIC) favoured linear relations over higher-order polynomials in all four cases. These $3\sigma$ envelopes were adopted as conservative estimates of the photospheric limits, minimizing contamination from sources with intrinsic infrared excess. The resulting relations are:

\begin{equation}
K-[\mathrm{IRAC}] = a + b\times [\mathrm{IRAC}]
\end{equation}
where the coefficients $(a,b)$ are:
$( -0.212,0.062)$, $(-0.511,0.095)$, $(-0.424,0.099)$, and
$(-0.782,0.139)$ for the [3.6], [4.5], [5.8], and [8.0] bands,
respectively.

At brighter magnitudes ($[\mathrm{IRAC}]<8.0$ mag), where the number of sources is insufficient to constrain the photospheric sequence, we adopted the values derived from the fitted relations at $[\mathrm{IRAC}] = 8.0$ (i.e., K–[3.6] = 0.287~mag, K–[4.5] = 0.248~mag, K–[5.8] = 0.367~mag, and K–[8.0] = 0.332~mag). For reference, Figure~\ref{fig:IRACclass} also shows the lower limits for CII objects from \citet{Luhman2010} for the colours K–[4.5] and K–[8.0] (0.7~mag and 1.15~mag, respectively), as well as the CII/CI boundaries corresponding to a flat SED between the K band and the IRAC bands (K–[3.6] = 1.49~mag, K–[4.5] = 2.22~mag, K–[5.8] = 2.97~mag, and K–[8.0] = 3.98~mag).

Based on their location in Figure~\ref{fig:IRACclass}, we define four IRAC flags in Table~\ref{tab:main} to identify different levels of infrared excess at 3.6 (flag\_3.6\micron), 4.5 (flag\_4.5\micron), 5.8 (flag\_5.8\micron), and 8.0~\micron\ (flag\_8.0\micron) for the main sample. Sources lying below the photospheric limits are classified as CIII. For the 3.6 and 4.5~\micron\ bands, sources located between the photospheric limits and the flat-SED boundaries are classified as CII, while those above the flat-SED boundaries are classified as CI. For flag\_5.8\micron\ and flag\_8.0\micron, sources located between the photospheric limits and the lower limits for CII objects from \citet{Luhman2010} 
are classified as evolved-disk (EV), following \citet{Hernandez2007}. Sources between the CII lower limits and the flat-SED boundaries are classified as CII, and those above the flat-SED boundaries are classified as CI.

\begin{figure}
    \centering
    \includegraphics[width=0.8\columnwidth]{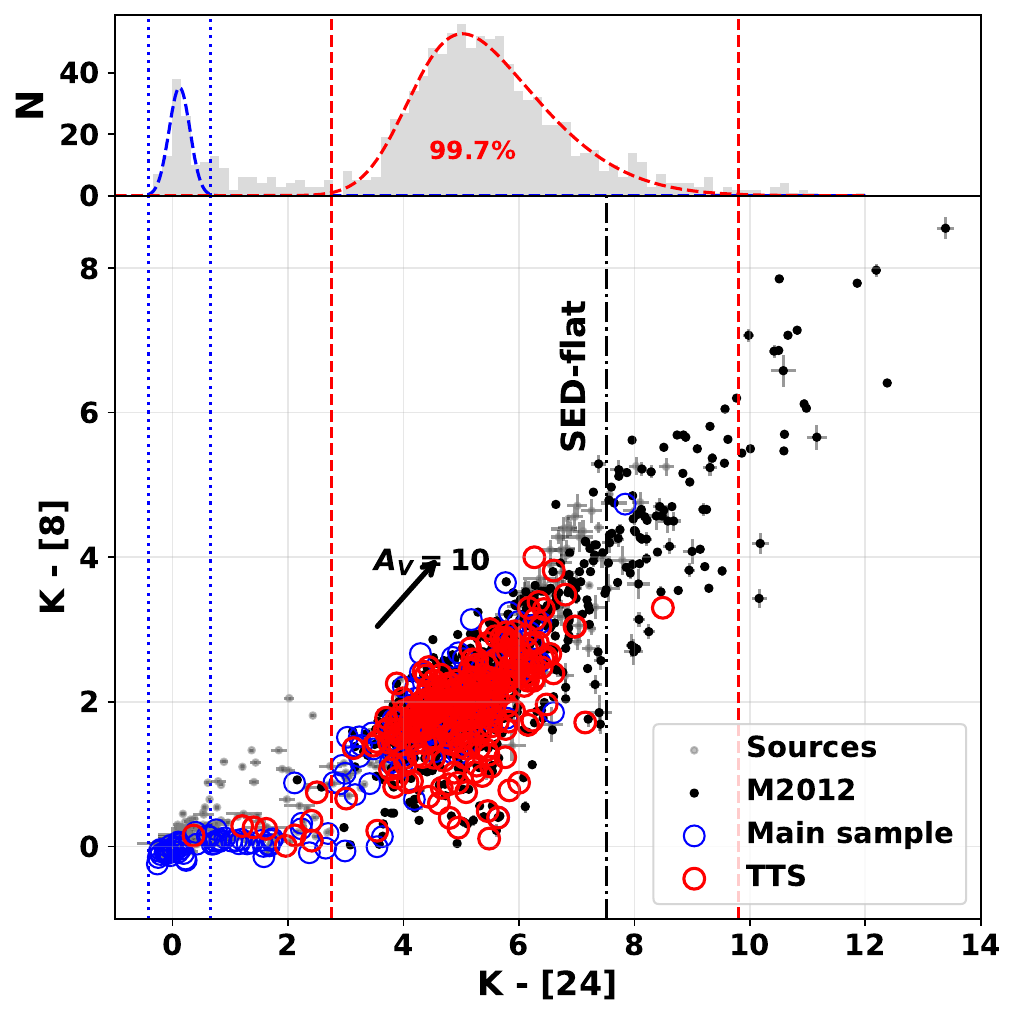}
    \caption{Colour-colour diagram used to identify stars with infrared excess at 24~\micron. The K-[24] colour is adopted due to its weak dependence on spectral type. The photospheric limit (blue line) is defined from a Gaussian fit to sources with K-[24] $\le$ 1~mag, adopting a 3$\sigma$ threshold (K-[24] = 0.66~mag). Given the asymmetric nature of the excess population, we characterize it using a skewed distribution with a median K-[24]$\sim$5.2~mag. The full-disc CII-regime is defined by the interval 2.76$<$K-[24]$<$9.8~mag, which encloses 99.7\% of the sources based on the cumulative distribution function. The vertical dot-dashed line indicates the SED-flat limit (K-[24] = 7.5~mag).}
    \label{fig:MIPSclass}
\end{figure}

Figure~\ref{fig:MIPSclass} shows the colour–colour diagram used to identify and classify stars with infrared excess at 24~\micron\ using MIPS photometry. We use the K–[24] colour, which is nearly independent of spectral type. The photospheric locus was characterized by selecting stars with K–[24] $\leq 1$~mag from the full infrared-detected sample and fitting a Gaussian function to their colour distribution, yielding a mean colour of K–[24] = 0.12~mag and a standard deviation of 0.18~mag. A conservative 3$\sigma$ threshold defines the upper envelope of the photospheric locus at K–[24] = 0.66~mag, minimizing contamination from sources with intrinsic infrared excess.

Because the distribution of the 24~\micron\ excess population (e.g., K–[24] $>$2~mag) is markedly asymmetric, we modelled it with a skew-normal profile rather than a Gaussian function. The distribution peaks at K–[24] $\sim$ 5.33~mag, and we define the CII regime within the interval 2.76 $<$ K–[24] $<$ 9.8~mag, which encloses 99.7\% of the sources based on the cumulative distribution function.

Based on the K–[24] colour, we defined a MIPS infrared flag to identify different levels of 24~\micron\ excess (flag\_24\micron) for the main sample corrected by reddening. Sources located blueward of the photospheric limit are classified as CIII, while those lying between the photospheric limit and the lower boundary of the CII regime are classified as EV. Objects falling between this boundary and the flat-SED limit are labelled as CII, and those red-ward of the flat-SED boundary are classified as CI.


\subsection{Theoretical Infrared Colours}
\label{s:colors}

To provide a theoretical reference for the infrared colours observed in the disc population, we computed synthetic colours for CII objects using the D’Alessio Irradiated Accretion Disk (DIAD) models \citep{DIAD1998,DIAD1999,DIAD2001,DIAD2006}. These models are intended to provide a global comparison with the observed colour distributions and characteristic SEDs, rather than a source-by-source characterization. The DIAD models self-consistently solve the equations of hydrostatic equilibrium and radiative energy balance to compute the radial and vertical disc structure, subsequently generating the SED for a disc with a specific set of physical parameters, and a viewing geometry described by $\cos(i)$. Here, $i$ is defined as the angle between the line of sight and the system's rotation axis. The heating sources include irradiation from the central star, accretion shocks on the stellar surface, and viscous dissipation within the disc.


The model input parameters include the stellar mass (\mstar), stellar radius (\rstar), \teff, mass accretion rate onto the stellar surface (\dotm), the \citet{Shakura1973} viscosity parameter ($\alpha$), the disc outer radius, and the parameter $z_{\mathrm{big}}$, which represents the height above the midplane of the layer containing large millimeter-sized grains. The models can be scaled using the distance to the system.
Dust is treated as two grain populations with 
different maximum sizes: small grains in the disc atmosphere with $a_{\mathrm{max,small}}$ = 0.25~\micron, and large grains concentrated toward the midplane with $a_{\mathrm{max,big}}$ = 1mm. 

The degree of dust settling is parameterized by $\epsilon = \zeta_{\mathrm{small}}/\zeta_{\mathrm{std}}$ \citep{DIAD2006}, where $\zeta_{\mathrm{small}}$ is the dust-to-gas mass ratio in the upper disc layers and $\zeta_{\mathrm{std}} = 0.01$ is the standard interstellar ratio value \citep{Drane1984}. 
For a given inclination $i$, 
the disc emission is computed using mean opacities 
\citep{DIAD1998,DIAD2006}. The radiative transfer equation is then solved along rays that sample the projected disc area toward the observer. The resulting SED includes contributions from the stellar photosphere, stellar light scattered by the disc surface, accretion shocks, the inner curved dust sublimation wall \citep[computed using a new self-consistent approach;][]{Manzo2026}, and the disc itself. The models also account for extinction produced by dust of the disc along the line of sight. 

\begin{figure}
    \centering
    \includegraphics[width=0.92\columnwidth]{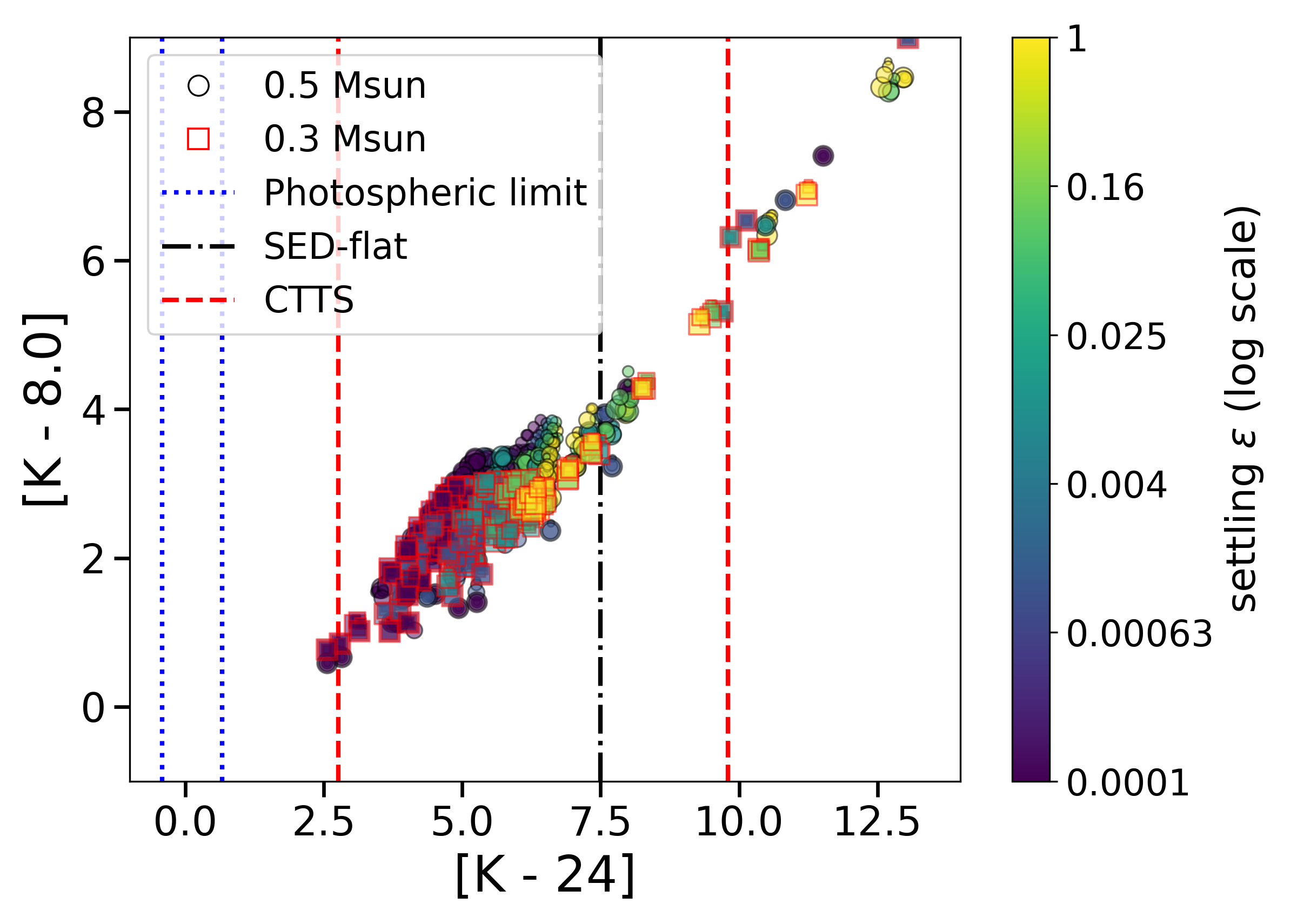}
    \caption{Synthetic colors for grids of disc models, for central stars with masses 0.3~\msun (squares) and a 0.5~\msun (circles). The colour bar indicates the $\epsilon$ settling parameter and the size of the symbol represents the value of $z_{\mathrm{big}}$. Vertical lines are the same as in Figure~\ref{fig:MIPSclass}.}
    \label{fig:daga2}
\end{figure}

The synthetic magnitudes were obtained by 
integrating the model SEDs over 
the corresponding 2MASS, IRAC, and MIPS filter response functions.
Figure~\ref{fig:daga2} shows the synthetic colours predicted by a grid of disc models computed by varying the mass accretion rate ($5\times10^{-10}\le\dot{M}\le1\times10^{-8}$ \msun\yrinv), the degree of dust settling ($0.0001\le\epsilon\le1$), the height of the millimeter-sized grain layer at the sublimation wall ($0.1H\le z_{\mathrm{big}}\le4H$, where $H$ is the gas scale-height at the wall location), and the inclination ($0.1\le\mathrm{\cos}(i)\le0.9$). For the central star, we adopted masses of 0.3 and 0.5 \msun. Models with higher degrees of settling (i.e., lower values of $\epsilon$) exhibit 
smaller infrared excesses above the photospheric colours, as expected. The theoretical locus spans the same region of the $K-[24]$ versus $K-[8.0]$ diagram occupied by the disc-bearing sources identified in Figure~\ref{fig:MIPSclass}, indicating that the observed infrared colours are broadly consistent with irradiated circumstellar discs spanning a range of accretion rates, dust-settling efficiencies, and viewing angles.


\subsection{Median empirical SEDs of CTTS in the ONC}
\label{s:SED}

\begin{figure}
    \centering
    \includegraphics[width=0.8\columnwidth]{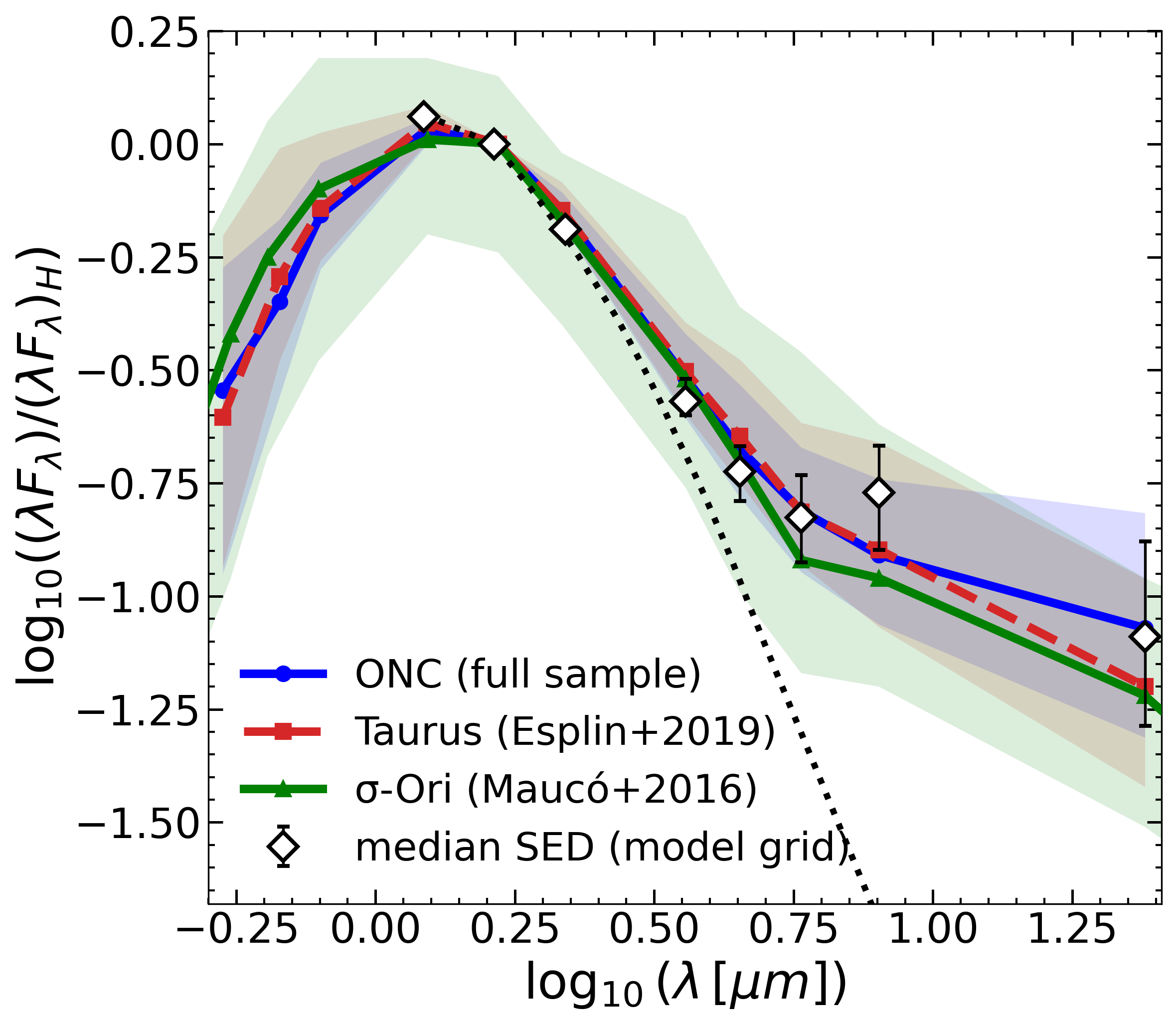}
    \caption{ Median SEDs of the ONC (blue), Taurus, from \citet{Esplin2019} (red), and $\sigma$~Ori from \citet{Mauco2016} (green). The shaded regions indicate the corresponding interquartile ranges. The diamonds and vertical bars represent the synthetic median SED and its interquartile range, respectively, obtained from a grid of disc models spanning a range of inclinations, dust-settling parameters, and mass accretion rates (see text for details). 
    The dotted line shows the photospheric SED of an M4 star from \citet{KH1995}, included as a visual reference to illustrate the infrared excesses.}
    \label{fig:medianONC}
\end{figure}

We constructed the median SED of CTTS in the ONC using \gaia 
(G, G$_{\mathrm{BP}}$, G$_{\mathrm{RP}}$), 2MASS (J, H, K), IRAC (3.6, 4.5, 5.8, and 8.0~\micron), and MIPS (24~\micron) photometry. 
The resulting H-band normalized median values are listed in Table~\ref{tab:median_sed}.
Figure~\ref{fig:medianONC} compares the ONC median SED with 
the median of Taurus, which we constructed using the sample from 
\citet{Esplin2019} and is listed in Table \ref{tab:median_Taurus_full} of Appendix \ref{app:medianSED}, and with the median SED of the $\sigma$~Ori cluster from \citet{Mauco2016}. All median SEDs were constructed for stars with spectral types between 
K0 and M6 and were normalized to the H-band. Unlike the analysis of Section~\ref{section:Massage}, which focused on the optical and near-infrared photospheric emission and therefore adopted the J-band as the reference, here we are primarily interested in the infrared excess at longer wavelengths. Thus, the H-band provides a more suitable anchor because it is less affected by extinction corrections than the J-band, while the K-band may already contain a significant contribution from disc emission \citep[e.g.,][]{DIAD1999, Kim2016, Manzo2026}. 

The synthetic median SED and the interquartile range from the grid of disc models described in Section~\ref{s:colors} (limiting the degree of dust settling to $0.0001 \leq \epsilon \leq 0.01$) are shown in Figure~\ref{fig:medianONC}.
Our synthetic grid 
covers a representative range of accretion rates and degrees of settling observed in discs \citep{Manara2016,Manzo2020}. 
The synthetic median, whose values are listed in Table~\ref{tab:median_sed_models}, reproduces the ONC median SED remarkably well, with the observed median lying within the interquartile range over the entire wavelength range.

The ONC and Taurus medians are in close agreement across 
the 2MASS 
(1.24--2.16~\micron) and 
the IRAC bands 
(3.6--8.0~\micron), while the Taurus median shows a lower flux at 24~\micron. A similar behaviour is observed for the $\sigma$~Ori median; however, in this case, the median SED lies below the ONC median from 5.8~\micron\ onward, and agrees well with the Taurus median at 24~\micron. 

The comparison between the ONC and Taurus medians for the full sample (Figure \ref{fig:medianONC}) appears to differ from the results of \citet{Kim2016}, who found that the ONC median lies above the Taurus median, particularly for stars with spectral types later than M2. However, \citet{Kim2016} constructed their median SEDs using spectra obtained with the Infrared Spectrograph \citep[IRS;][]{Houck2004} onboard the \spitzer Space Telescope,
which may introduce a selection bias toward brighter discs. In contrast, our analysis is based on photometry, for which such a bias is expected to be less severe. 

\begin{figure}
    \centering
    \includegraphics[width=0.7\columnwidth]
    {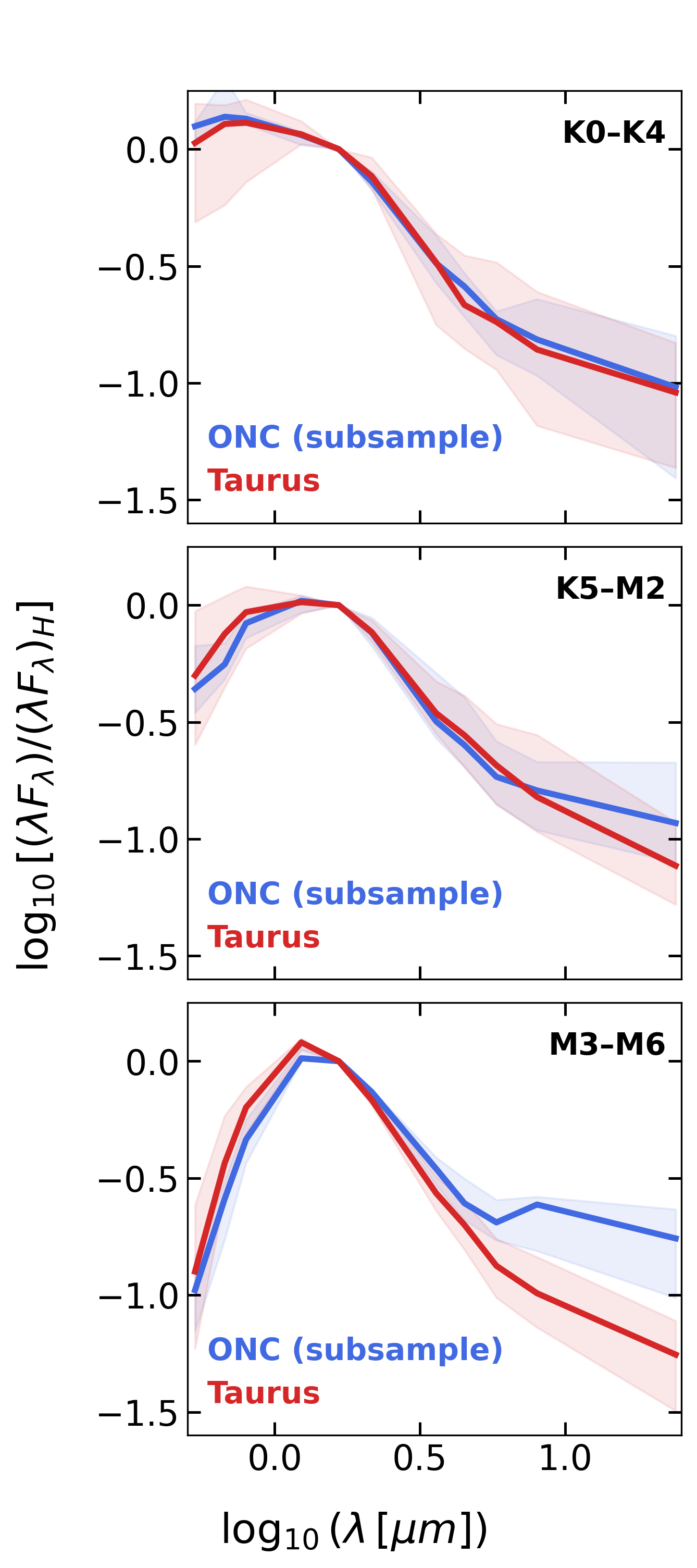}
    \caption{ Median SEDs separated by spectral types bins, as indicated by the labels. The ONC medians (blue) were obtained from the subsample resulting from the cross-match between our CTTS sample and that of \citet{Kim2016}, while the Taurus median SED from \citet{Esplin2019} is shown in red. The shaded regions indicate the corresponding interquartile ranges. In particular, the subsample of stars later than M2 reproduces the trend reported by \citet{Kim2016}, with the ONC median SED lying above the Taurus median.}
    \label{fig:medianONC_sptyp}
\end{figure}

To investigate this possibility, we cross-matched our ONC CTTS sample with that of 
\citet{Kim2016} and reconstructed the ONC median SED for this cross-matched subsample (hereafter the ONC-Kim subsample), 
separating the stars into similar spectral type bins as the ones adopted by 
\citet{Kim2016}: K0--K4, K5--M2, and M3--M6. For this restricted 
subsample, we recover a trend similar to that reported by 
\citet{Kim2016}, as shown in Figure~\ref{fig:medianONC_sptyp}. 
In particular, the K5–M2 stars exhibit ONC and Taurus median SEDs that agree within the interquartile ranges throughout the IRAC wavelength range, whereas the M3–M6 subsample shows an ONC median that lies systematically above the Taurus median. 
Table \ref{tab:median_Taurus_ONC_SpT} in Appendix \ref{app:medianSED} lists the median SEDs for Taurus and for the ONC-Kim subsample, both separated by spectral type.

In addition, we compared the distributions of the IRAC 
[5.8] and [8.0] magnitudes for both our full ONC CTTS sample and the
subsample cross-matched with \citet{Kim2016}. The median [5.8] magnitudes 
are 9.77~mag and 8.86~mag for the full sample and the cross-matched subsample, 
respectively. Similarly, the median [8.0] magnitudes are 8.97~mag and 8.08~mag, 
respectively. These differences indicate that the \citet{Kim2016} subsample 
is biased toward brighter mid-infrared sources, which likely explains the apparent discrepancy between their results and the median SED obtained from our full photometric sample of K0–M6 stars. 

Overall, for the full K0–M6 sample, Figure~\ref{fig:medianONC} may suggest that discs in $\sigma$~Ori are at a more advanced evolutionary stage than those in the ONC and Taurus, particularly in the inner regions($\lesssim$10~au), from which most of the emission at wavelengths $\lesssim$ 24~\micron\ arises \citep[][]{DIAD2006, Manzo2026}. It is important to note, however, that this interpretation cannot be drawn from the median SEDs alone, since differences in sample selection between the regions may also affect the comparison. For example, as discussed above, the use of Spitzer/IRS spectroscopy to construct median SEDs can introduce a bias toward brighter discs \citep{Kim2016}. In contrast, the $\sigma$~Ori disk census is based on photometry, which reaches fainter sources and is therefore less susceptible to this specific selection bias. Nevertheless, independent evidence for disk evolution with age has been obtained from the characterization of disk properties in stellar populations of different ages, including the decrease in disk dust masses and the increase in the degree of dust settling with age \citep[e.g.][]{Grant_2018, Manzo2020, Rilinger_2023}.

In addition, the similarity between the ONC and Taurus median SEDs does not reveal a clear difference in the evolution of the inner disc regions ($\lesssim 10$~au) between the two populations. Although environmental effects associated with massive stars and higher stellar densities in the ONC may influence disc evolution, assessing their contribution would require a dedicated analysis that accounts for the different properties of the two populations. External photoevaporation is expected to primarily affect the outer disc, particularly for discs located close to massive stars \citep[e.g.,][]{Balog2007,Fang2012,Mauco2016}, but its impact is not quantified in the present here, as this is beyond the scope of the present study.


\begin{table}
\centering
\caption{Median H-normalized SED and quartiles for the full ONC sample.}
\begin{tabular}{c c c c}
\hline
Wavelength [\micron] & 
$\log_{10}\left[(\lambda F_\lambda)/(\lambda F_\lambda)_H\right]$ 
& Lower & Upper  \\
\hline
\hline
0.530 & -0.545 & -0.950 & -0.273 \\
0.640 & -0.348 & -0.557 & -0.166 \\
0.780 & -0.157 & -0.276 & -0.041 \\
1.22 &  0.032 & -0.003 &  0.057 \\
1.63 &  0.000 &  0.000 &  0.000 \\
2.19 & -0.148 & -0.177 & -0.106 \\
3.60 & -0.517 & -0.606 & -0.420 \\
4.50 & -0.674 & -0.781 & -0.532 \\
5.80 & -0.811 & -0.946 & -0.671 \\
8.00 & -0.909 & -1.062 & -0.741 \\
24.0 & -1.070 & -1.312 & -0.816 \\
\hline
\hline
\end{tabular}
\label{tab:median_sed}
\end{table}

\begin{table}
\centering
\caption{Theoretical median H-normalized SED and quartiles for the model grid.}
\begin{tabular}{c c c c}
\hline
Wavelength [\micron] & 
$\log_{10}\left[(\lambda F_\lambda)/(\lambda F_\lambda)_H\right]$ 
& Lower & Upper  \\
\hline
\hline
1.22  &  0.060 &  0.046 &  0.066 \\
1.63  &  0.000 &  0.000 &  0.000 \\
2.19  & -0.189 & -0.194 & -0.171 \\
3.60  & -0.569 & -0.600 & -0.520 \\
4.50  & -0.724 & -0.790 & -0.668 \\
5.80  & -0.825 & -0.926 & -0.732 \\
8.00  & -0.771 & -0.898 & -0.667 \\
24.0  & -1.089 & -1.287 & -0.879 \\
\hline
\hline
\end{tabular}
\label{tab:median_sed_models}
\end{table}

\subsection{Estimates of mass accretion rate}
\label{s:acretion}

\begin{figure}
    \centering
    \includegraphics[width=0.85\columnwidth]{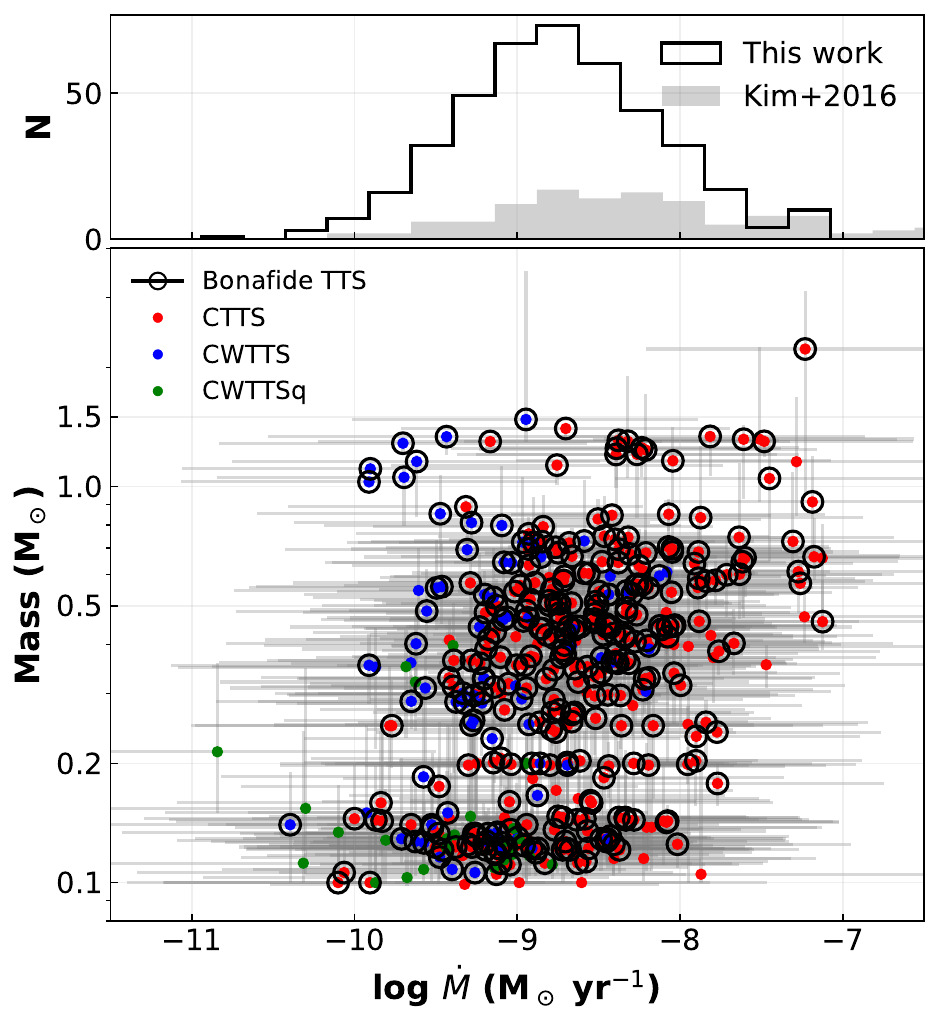}
    \caption{Bottom panel: mass accretion rate versus stellar mass \citep[derived from \massage, adopting the spotless models of ][]{Somers2020} for accreting TTS with spec\_flag = CTTS (red), CWTTS (blue), and CWTTSq (green). Bona fide TTS are highlighted with black open circles. Gray error bars correspond to the 16th–84th percentile intervals obtained from the MC propagation of uncertainties. Top panel: comparison between the distribution of $\log $\dotm derived in this work (black histogram) and that reported by \citep[gray histogram]{Kim2016}.}
    \label{fig:accretion}
\end{figure}

To compute the mass accretion rates (\dotm) for the ONC sample, we extended the methodology described in \citet{Manzo2020} by implementing an MC error propagation approach \citep{Anderson1976}. We performed 10$^5$ MC iterations to ensure numerical convergence of the median and higher-order moments of $\mathrm{L}_{\mathrm{H}\alpha}$, $\mathrm{L}_{\mathrm{acc}}$, and \dotm. 
The MC framework incorporates uncertainties from the \gaia observations (G$_\mathrm{BP}$, G$_\mathrm{RP}$, $\Pi_{corr}$), as well as from the stellar parameters derived with the \massage code (Section~\ref{section:Massage}; \mstar, L$_\ast$, \teff, \av). In the absence of formal uncertainties, we adopted a conservative uncertainty of 10\% for the EW$_{\mathrm{H}\alpha}$ measurements obtained from \sptclass (Section~\ref{sec:sptclass}).
The mass accretion rate was computed as follows:

\begin{enumerate}

\item We derived the \Halpha line luminosity $L_{\mathrm{H}\alpha}$ from EW$_{\mathrm{H}\alpha}$.
The continuum flux ($F_{\mathrm{c}}$) at \Halpha was estimated via a linear interpolation between the dereddened \gaia fluxes in the G$_\mathrm{BP}$ and G$_\mathrm{RP}$ bands.
Zero-point fluxes and effective filter wavelengths for these bands were obtained from the Spanish Virtual Observatory (SVO) Filter Profile Service\footnote{\url{https://svo2.cab.inta-csic.es/theory/fps/}}. The line luminosity was then computed as:
\begin{equation}
L_{\mathrm{H}\alpha}=4\pi\times F_{\mathrm{c}} \times \mathrm{EW}_{\mathrm{H}\alpha}\times(1000/\Pi_{\mathrm{corr}})^2
\end{equation}
The MC realizations yield a probability density function (PDF) for L$_{\mathrm{H}\alpha}$.

\item The accretion luminosity (L$_{\mathrm{acc}}$) was estimated using an empirical correlation with L$_{\mathrm{ H}\alpha}$ of the form:
\begin{equation}
\log(\mathrm{L}_{\mathrm{acc}})=a(\pm\sigma_a) \times \log(\mathrm{L}_{\mathrm{H}\alpha})+b(\pm\sigma_b).
\end{equation}
To propagate the uncertainties of the empirical calibration, the coefficients $a$ and $b$ were randomly sampled according to their reported uncertainties at each MC iteration.
The resulting L$_{\mathrm{acc}}$ values were also characterized by a PDF. 
We adopted the empirical correlations from \citet{Ingleby13} (where $a=1.0$, $\sigma_a = 0.2$, $b = 1.3$, $\sigma_b  =0.7$) and from \citet{Alcala14} (where $a=1.12$, $\sigma_a=0.07$, $b=1.5$, $\sigma_b=0.26$). Both correlations are calibrated for low-mass stars ($\lesssim$1.5\msun), consistent with the mass range of most TTS in our sample. 
Since the coefficients of the two calibrations are similar, the resulting mass accretion rates differ by less than $\sim$0.25~dex.

\item Finally, we derived the mass accretion rate from the accretion luminosity using the magnetospheric accretion formalism of \cite{Gullbring98},
\begin{equation}
\label{ec4}
\dot{\mathrm{M}}=\biggl(1-\frac{\mathrm{R}_\ast}{\mathrm{R}_{\mathrm{in}}}\biggr)^{-1}\frac{\mathrm{L}_{\mathrm{acc}} \mathrm{R}_\ast}{G\mathrm{M}_\ast}.
\end{equation}

The magnetospheric truncation radius $\mathrm{R}_{\mathrm{in}}$, in units of the stellar radius, was fixed at 5\rstar \citep{Hartmann2016}.
The stellar radius \rstar was derived self-consistently from the randomly sampled distributions of the stellar luminosity L$_\ast$ and \teff from \massage. We adopted the stellar mass derived from the spotless models of \cite{Somers2020}, listed in the mass\_S00 column of Table 3.
Because \mstar enters explicitly into Equation \ref{ec4}, the Monte Carlo realizations of \mstar and  \dotm are formally anti-correlated. We find this correlation to be very weak, with a median Pearson coefficient of r=-0.082.
\end{enumerate}

Table~\ref{tab:main} lists the median values of L$_{\mathrm{ acc}}$ and \dotm, along with the 16th and 84th percentiles, which encompass approximately 68\% of the MC-derived distributions. These values were computed for accreting TTS, defined as sources with \texttt{spec\_flag} = CTTS, CWTTS, or CWTTSq.

Figure~\ref{fig:accretion} shows the relation between \dotm and \mstar for accreting TTS, highlighting the bona fide TTS sample. The large scatter in the derived accretion rates prevents us from identifying a strong dependence of \dotm on stellar mass within our sample. For the bona fide CTTS, we obtain a correlation coefficient of approximately 0.46 between $\log$\mstar and $\log$\dotm, similar to that obtained from the ONC sample analysed by \citet{Kim2016}. 
Because \mstar enters explicitly into Equation \ref{ec4}, it introduces an inverse dependence of \dotm on \mstar for fixed $\mathrm{L}_{\mathrm{acc}}$ and \rstar. However, $\mathrm{L}_{\mathrm{acc}}$ and \rstar
also vary across our sample and may contribute to the observed positive correlation. For the full sample of accreting TTS, the correlation coefficient decreases to below 0.40. This contrasts with studies of other star-forming regions, where a stronger correlation between stellar mass and accretion rate has been reported \citep[e.g.,][]{Alcala14}. 
The origin of this difference remains unclear. Possible explanations include differences in sample selection, the relatively narrow stellar mass range covered by our survey, the adopted accretion diagnostics and methodologies used to derive accretion rates, and the intrinsic variability of accretion, all of which can contribute to the observed scatter.

In the upper panel of Figure~\ref{fig:accretion}, we compare the \dotm distributions derived in this work with those reported by \citet{Kim2016}, who estimated accretion rates using the hydrogen recombination lines Pa$\gamma$ (1.094~\micron), Pa$\beta$ (1.282~\micron), and Br$\gamma$ (2.166~\micron). The distribution reported by \citet{Kim2016} is shifted toward higher accretion rates.
The median values of $\log(\dot{M}/M_\odot)$ are -8.34 (Q1–Q3: -8.81 to -7.83) and -8.75 (Q1–Q3: -9.17 to -8.38) for the samples of \citet{Kim2016} and this work, respectively.

As discussed in Section~\ref{s:SED}, the sample analysed by \citet{Kim2016} is biased toward brighter mid-infrared sources, which are also expected to exhibit, on average, higher accretion rates. This selection effect likely contributes to the observed offset between the two distributions. The difference may also be partially related to systematic differences between accretion diagnostics. Previous studies have shown that accretion rates derived from H$\alpha$ can differ from those obtained using hydrogen recombination lines at longer wavelengths \citep[e.g.,][]{Alcala14}. The \Halpha line is more sensitive to chromospheric emission, variability, and extinction uncertainties than near-infrared accretion tracers, which may contribute to the larger dispersion observed in our \dotm estimates. Additionally, recent multicolumn magnetospheric accretion models further suggest that Balmer line emission may arise from the superposition of accretion flows with different geometries and accretion rates, rather than from a single homogeneous accretion column \citep{Patino2026}. Such complexity may introduce additional scatter in the empirical L$_{\mathrm{H}\alpha}$-L$_{\mathrm{acc}}$ relation used to estimate mass accretion rates.

\section{Summary and Conclusions}
\label{sec:Conclusions}

In this work, we present a comprehensive and homogeneous characterization of the stellar population in the ONC by combining low-resolution optical spectroscopy, \gaia data, and infrared photometry. 
Using a homogeneous analysis framework, we derived spectral classifications, membership constraints, stellar parameters, disc diagnostics, and mass accretion rates for a large sample of stars in the cluster. Our main results can be summarized as follows:

\begin{itemize}
\item 
We present a spectroscopic census of 1,513 stars in the ONC region based primarily on MMT/Hectospec observations, complemented with spectra from SPMO, MDMO, and FLWO. From the analysis of \ion{Li}{I} and \Halpha, combined with \gaia astrometry, we identify 780 bona fide TTS, together with 40 IMTTS and 13 HAeBe stars.

\item We combine spectroscopic diagnostics with precise astrometry to identify 2804 kinematic PMS candidates in the ONC region,
enabling a refined characterization of the cluster structure and helping disentangle overlapping stellar populations.


\item We derive homogeneous extinctions, luminosities, masses, ages, and surface gravities using the \massage code, developed for the characterization of young stellar populations. Comparisons among evolutionary models reveal systematic differences in the inferred stellar parameters, highlighting the importance of model assumptions in PMS studies.

\item We revisit the IRAC and MIPS \spitzer photometry to construct an updated census of protoplanetary discs. The median infrared SED of ONC stars is remarkably similar to that of Taurus, suggesting that environmental effects associated with the high stellar density and massive stars do not significantly accelerate the overall evolution of the inner disc regions ($\lesssim$ 10~au). In contrast, the median SED of the slightly older $\sigma$~Ori cluster indicates a more evolved disc population.

\item We estimate mass accretion rates using \Halpha diagnostics and MC error propagation approach. The distribution of accretion rates spans approximately -10 $\lesssim \log(\dot{M}) [M_{\odot} yr^{-1}] \lesssim$-7, while the large intrinsic scatter prevents the identification of a strong dependence between mass accretion rate and stellar mass, consistent with previous studies of the ONC.

\item The homogeneous catalogue presented in this work combines spectroscopic classifications, membership information, infrared diagnostics, and self-consistent stellar parameters for a large sample of ONC members. This dataset provides a valuable reference for future studies aimed at investigating the star formation history, kinematic substructure, and disc evolution of the ONC.
\end{itemize}

Overall, this work provides a unified view of the ONC stellar population by linking spectroscopic properties, kinematics, stellar parameters, circumstellar discs, and accretion diagnostics within a homogeneous framework, establishing a robust observational benchmark for studies of young stellar populations in the \gaiam era.

\section*{Acknowledgements}
J.H. acknowledges support from UNAM--DGAPA--PAPIIT through grant IN110126. E.M.M. and L.A. acknowledge support from the Secretaría de Ciencia, Humanidades, Tecnología e Innovación (SECIHTI) through postdoctoral fellowships under the ``Estancias Posdoctorales por México'' program. S.S.-S., L.M., and C.Z. acknowledge financial support from the National Postgraduate Scholarship Program of SECIHTI, M\'exico. N.C. and E.M.M. acknowledge support from NASA XRP grant 80NSSC22K0151. C.R.-Z. acknowledges support from UNAM--DGAPA--PAPIIT through grant IN107226. We acknowledge the Activity and Rotation of Young Stellar Objects (ARYSO) and the Models and Observation of Disk Evolution in Latin America (MODELA) collaboration groups for valuable discussions that helped improve this work.

This work is based in part on observations obtained with the Hectospec multi-object spectrograph on the 6.5 m MMT Observatory, a joint facility of the Smithsonian Institution and the University of Arizona. This work is also based upon observations carried out at the Observatorio Astronómico Nacional on the Sierra San Pedro Mártir (OAN-SPM), Baja California, M\'exico. We thank the daytime and night support staff at the OAN-SPM for facilitating and helping obtain our observations. Additional observations were obtained at the MDM Observatory, operated by the University of Michigan, Dartmouth College, Ohio State University, Columbia University, and Ohio University, and with the FAst Spectrograph (FAST) on the 1.5 m Tillinghast Telescope at the Fred Lawrence Whipple Observatory (FLWO).

This work has made use of data from the European Space Agency (ESA) mission \gaiam, processed by the \gaiam Data Processing and Analysis Consortium (DPAC). This publication also makes use of data products from the Two Micron All Sky Survey (2MASS), a joint project of the University of Massachusetts and IPAC/Caltech, funded by NASA and the National Science Foundation. In addition, this work is based in part on observations made with the \spitzer Space Telescope and has made use of the NASA/IPAC Infrared Science Archive (IRSA), which is funded by NASA and operated by the California Institute of Technology. This research has made use of the SIMBAD database and the VizieR catalogue access tool, operated at CDS, Strasbourg, France. This research has also made use of NASA's Astrophysics Data System Bibliographic Services and the SVO project funded by MCIN/AEI/10.13039/501100011033 through grant PID2023-146210NB-I00.

\section*{Facilities}
MMT: 6.5m (Hectospec), OAN-SPM: 2.1m (Boller \& Chivens), MDM:1.3m (CCDS), FLWO:1.5m (FAST), \gaia, \spitzer (IRAC, MIPS), 2MASS.

\section*{Software}

Python, IDL, IRAF, TOPCAT \citep{Taylor2005}, \sptclass
\citep{Hernandez2004,Hernandez2017}, DIAD \citep{DIAD1998,DIAD1999,DIAD2001,DIAD2006},
and \massage (this work; available at \url{https://github.com/jesush149/MassAge_py}).

\section*{Data Availability}


The complete versions of Tables \ref{tab:main}, \ref{tab:mass_and_ages}, \ref{tab:mass_and_ages2}, and \ref{tab:kinmem_additional} are available as online supplementary material associated with this article, and will also be available through the CDS/VizieR service upon publication.



\bibliographystyle{mnras}
\bibliography{example} 




\appendix
\section{Kinematic members not included in the present study}
\label{ap:kinmem_teff}
In this appendix, we compile complementary information for a sample of kinematic candidates that are not included in the main spectroscopic analysis presented in this work. These sources are identified as potential members of the ONC based on their astrometric properties from \gaia, but lack optical spectra in our dataset and therefore were not included in the derivation of spectral types, stellar parameters, and accretion properties described in the main body of the paper.

For these objects, we collected spectral types and \teff from previous studies available in the literature. When only \teff estimates are reported, we convert them into spectral types using the standard calibrations from \citet{Pecaut2013} to ensure consistency across the compiled sample.

In addition, we include infrared photometry from the \spitzer Space Telescope to analyse the infrared properties of this sample. Infrared excesses are identified using the [5.8], [8.0], and [24]~\micron\ bands, following the criteria described in Section~\ref{sec:infrared}. 

Table~\ref{tab:kinmem_additional} provides an extended view of the kinematically selected population in the ONC region, allowing us to assess their basic stellar properties and disc incidence, and to place them in the broader context of the cluster population analysed in this work.

\begin{table}
\centering
\caption{Additional kinematic candidates}
\begin{tabular}{ll}
\hline
Column label & Description       \\
\hline \hline
Gaia DR3\_ID   &  \gaia source name \\
SpT & spectral types \\
REF\_SpT & reference for spectral types \\
$[3.6]$ & \spitzer photometry, 3.6 \micron \\
e\_$[3.6]$ & error of the \spitzer photometry, 3.6 \micron \\
$[4.5]$ & \spitzer photometry, 4.5 \micron \\
e\_$[4.5]$ & error of the \spitzer photometry, 4.5 \micron \\
$[5.8]$ & \spitzer photometry, 5.8 \micron \\
e\_$[5.8]$ & error of the \spitzer photometry, 5.8 \micron \\
$[8.0]$ & \spitzer photometry, 8.0 \micron \\
e\_$[8.0]$ & error of the \spitzer photometry, 8.0 \micron \\
$[24]$ & \spitzer photometry, 24.0 \micron \\
e\_$[24]$ & error of the \spitzer photometry, 24.0 \micron \\
REFERENCE & \citet{Megeath2012} or this work \\
flag\_3.6\micron & Infrared analysis: CIII, CII, CI \\
flag\_4.5\micron & Infrared analysis: CIII, CII, CI \\
flag\_5.8\micron & Infrared analysis: CIII, EV, CII, CI \\
flag\_8.0\micron & Infrared analysis: CIII, EV, CII, CI \\
flag\_24\micron & Infrared analysis: CIII, EV, CII, CI \\
Comments & comments \\ 
\hline
\end{tabular}
\label{tab:kinmem_additional}
\end{table}

\section{Description of the \massage Code and User Options}
\label{ap:massage_procedure}
Using as input the \gaia photometry, the J and H 2MASS magnitudes, parallaxes, and spectral types (or \teff), the Python-based \massage code estimates {\av}, luminosities, ages, masses, and \logg. The uncertainties in the outputs are estimated using the MC propagation method \citep{Anderson1976}, assuming that the input uncertainties follow Gaussian distributions. The user can choose the number of MC iterations, $nmc$. The \massage procedures and user options are described in Sections~\ref{ap:SpT_Teff}, \ref{app:Lum}, and~\ref{ap:massage} and in Table~\ref{t:massage}, while the caveats and limitations of \massage are described in Section~\ref{app:caveats}.

\subsection{Spectral type-\teff conversion and visual extinction determination.}
\label{ap:SpT_Teff}

When spectral types are provided as input, we convert them to \teff using standard calibrations \citep[e.g.,][]{KH1995,Pecaut2013}. Alternatively, \teff and its associated uncertainties can be directly provided in either a linear or a logarithmic scale. Standard colours and the bolometric corrections (BC) can be selected from empirical or theoretical calibrations (see Table~\ref{t:massage}).

For the estimation of \av, the user must select an extinction law \citep[e.g.,][]{CCM1989,Schlafly2016,Fitzpatrick2019} and a value of the total-to-selective extinction ratio, $R_V$. A value of $R_V\approx$ 3.1 is typical of the diffuse interstellar medium. Larger values ($R_V\sim$ 4–5) imply a flatter extinction curve and are generally associated with denser environments and larger average dust grain sizes \citep[e.g. molecular clouds and protoplanetary discs;][]{Hernandez2004}.

By default, \massage adopts the nominal filter wavelengths in the extinction law to compute the extinction in each photometric band \al ($A_{G_\mathrm{BP}}/A_V$, $A_{G}/A_V$, $A_{G_\mathrm{RP}}/A_V$, $A_J/A_V$, $A_H/A_V$). The visual extinction is derived by minimizing the differences between the observed colours corrected for extinction,
\begin{equation}
    [M_\lambda-M_0]_{corr}=[M_\lambda-M_0]_{obs}-\biggl(\frac{A_\lambda}{A_V}-\frac{A_0}{A_V}\biggr)\times A_{V}
\end{equation}
and the expected standard colours $[M_\lambda-M_0]_{std}$ for a given \teff, varying \av according to:

\begin{equation}\label{e:extinction}
D[A_V]=\sqrt{\frac{\sum([M_\lambda-M_0]_{corr} - [M_\lambda-M_0]_{std})^2}{n}},
\end{equation}
where $M_0$ denotes the anchor magnitude, in which the standard magnitude ([$M_0$]$_{std}$) is equal to the extinction corrected observed magnitude ([$M_0$]$_{obs}$-$A_0/A_V$). The anchor magnitude can be J (default option), $G_\mathrm{RP}$, or $G$. $M_\lambda$ represents the remaining magnitudes, and $n$ represents the number of colours used in the minimization.

\massage also provides an alternative option in which the effective wavelength $\lambda_{eff}$ of each photometric filter is computed according to the expected stellar SED \citep{Bessell2012}. For certain sources (e.g., highly embedded low-mass stars), the effective wavelength may differ significantly from the nominal filter wavelength, which can produce measurable differences in the derived extinction. In this approach, the effective wavelength is obtained by 
integrating 
 the stellar spectrum ($S_\lambda$[AV])
weighted by
the filter transmission curve ($T_\lambda$) following the equation: 

\begin{equation}
\label{e:WLeff}
\lambda_{eff}=\frac{\int{\lambda\times T_\lambda\times S_\lambda[A_V]\times d\lambda}}{{\int{T_\lambda\times S_\lambda[A_V]\times d\lambda}}}.
\end{equation}
$S_\lambda$[AV] is estimated from a low-resolution (R = 50) theoretical PHOENIX stellar atmosphere models spectra \citep{Husser2013} with \teff similar to the target, 
 to which an extinction corresponding to {\av} is applied.

\begin{table*}
\centering
\caption{\massage configuration options}\label{t:massage}
\begin{tabular}{lll}
\hline
Parameter & options & Description       \\
\hline \hline
X axis & Teff     & input in linear scale\\
       & logTeff  & input in logarithmic scale\\
       & SpT\_PM13& use the \citet{Pecaut2013} conversion table\\
       & SpT\_KH95& use the \citet{KH1995} conversion table\\
Y axis & PM13   & Standard colours from \citet{Luhman2020} and BC from \citet{Pecaut2013}\\
       & KH95   & Standard colours from \citet{Luhman2020} and BC from \citet{KH1995}\\
       & PARSEC & theoretical values from PARSEC models\\
       & MIST & theoretical values from MIST models\\ 
Extinction law & c89   &  \cite{CCM1989}\\
                & s16   &  \cite{Schlafly2016}\\
                & f19   &  \cite{Fitzpatrick2019}\\
$R_V$          & float   &  the total-to-selective extinction ratio, typically between 3 and 5 \\
Anchor filter & J, G$_\mathrm{RP}$, G  &  Standard magnitude = extinction corrected magnitude \\
Model grid & PARSEC  &  \citet{Marigo2017}: 0.01$<$age$<$1000 Myr; 0.09$<$mass$<$30 \msun \\
           & MIST  &  \citet{Dotter2016}: 0.001$<$age$<$1000 Myr; 0.1$<$mass$<$30 \msun \\
           & SF00  &  \citet{Siess2000}: 0.001$<$age$<$22000 Myr; 0.1$<$mass$<$7 \msun \\
           & PISA  &  \citet{Tognelli2011}: 0.008$<$age$<$62000 Myr; 0.2$<$mass$<$7 \msun\\
           & Baraffe  &  \citet{Baraffe2002,Baraffe2015}: 1.0$<$age$<$10000 Myr; 0.001$<$mass$<$1.5 \msun \\
           & Feiden\_std  &  Standard Model \citep{Feiden2016}: 0.1$<$age$<$1000 Myr; 0.08$<$mass$<$3.0 \msun \\
           & Feiden\_mag  &  Magnetic Model \citep{Feiden2016}: 0.1$<$age$<$1000 Myr; 0.09$<$mass$<$2.45 \msun \\
           & Somers\_0  &  no spot \citep{Somers2020}: 0.15$<$age$<$15000 Myr; 0.1$<$mass$<$1.3 \msun \\
           & Somers\_17  & spot filling factor 0.17 \citep{Somers2020}: 0.15$<$age$<$15000 Myr; 0.1$<$mass$<$1.3 \msun \\
           & Somers\_34  & spot filling factor 0.34 \citep{Somers2020}: 0.15$<$age$<$15000 Myr; 0.1$<$mass$<$1.3 \msun  \\
           & Somers\_51  & spot filling factor 0.51 \citep{Somers2020}: 0.15$<$age$<$15000 Myr; 0.1$<$mass$<$1.3 \msun   \\
           & Somers\_68  & spot filling factor 0.68 \citep{Somers2020}: 0.15$<$age$<$15000 Myr; 0.1$<$mass$<$1.3 \msun   \\
           & Somers\_85  & spot filling factor 0.85 \citep{Somers2020}: 0.15$<$age$<$15000 Myr; 0.1$<$mass$<$1.3 \msun   \\
Evolutionary phase & pms & Pre-Main-Sequence phase, only for PARSEC and MIST\\
                 & ms  & Pre-Main-Sequence and Main-Sequence phases, only for PARSEC and MIST\\
                 & all & no phase restriction\\
interpolation mode$^{(1)}$ & linear & interpolation based on the Delaunay triangulation\\
                   & cubic  & Uses a piecewise cubic, continuously differentiable surface. \\
                   & nearest-neighbour & Uses the value of the nearest data point. \\
Min mass and Max mass & \msun & minimal and maximal masses used from the model\\
Min age and Max age & Myr & minimal and maximal ages used from the model\\
Monte Carlo iterations & integer &number of iteration for the MC error propagation method \\ 
\hline
\end{tabular}
\begin{tablenotes}
\item (1) The two-dimensional interpolation uses the interpolate.griddata function from the SciPy library.
\item Users can activate the option to estimate extinction by calculating effective filter wavelengths instead of nominal values.  
\end{tablenotes}
\end{table*}
\subsection{Bolometric luminosity estimation}
\label{app:Lum}

Following a standard procedure, we estimate the absolute bolometric magnitude for the anchor filter using the apparent magnitude corrected by extinction, the parallax, and the bolometric correction ($M_{abs}$ = $M_0-5\log(1000/\Pi_{corr})+5-A_0+BC$). Then, we applied the equation:

\begin{equation}
\label{e:Lum}
\log\left(\frac{L}{L_\odot}\right)=\frac{4.7554-M_{bol}}{2.5},
\end{equation}
where 4.7554 is the solar absolute bolometric magnitude \citep{Mamajek2012}

\subsection{Inference of mass, age, and surface gravity from evolutionary models}
\label{ap:massage}

Stellar masses and ages are determined by comparing the positions of each source on the HRD with theoretical stellar evolutionary models. These models assume solar metallicity, and the user can set the maximum and minimum values of age and mass to mitigate degenerate solutions. Also, for the MIST and PARSEC stellar evolutionary models, the user can select specific evolutionary phases (e.g., PMS, MS). For each interpolated point, theoretical values of mass, age, and \logg are obtained using two-dimensional interpolation (linear, cubic, or nearest-neighbour) on the \teff-\logLs theoretical plane.

Since \massage uses an MC error-propagation method, it generates $nmc$ possible results. For each output parameter, the resulting table includes the mean and standard deviation, the median and median absolute deviation, the lower and upper limits, and the skewness of the resulting distribution. 
The lower and upper limits contain 68.2\% of the resulting distribution, which is equivalent to the 1$\sigma$ interval of a normal distribution. Finally, when possible results are interpolated beyond the theoretical limits of the selected model, a specific flag is included in the output.

\subsection{Caveats and Limitations}
\label{app:caveats}
\begin{itemize}

\item All targets are treated as single stars in this analysis. However, unresolved binary or multiple systems can bias the derived parameters, as the combined flux leads to overestimated luminosities, systematically inferred younger ages, and higher inferred masses in the HRD. The reported values should therefore be interpreted with caution, particularly when multiplicity constraints are not available.

\item The derived stellar masses and ages are model-dependent. Models that include surface spot coverage or magnetic inhibition of convection predict stellar structures that differ from those of standard (non-magnetic, unspotted) models, leading to systematic differences in the inferred parameters when adopting the same observed \teff and \logLs (see Figures \ref{fig:mass1}, \ref{fig:mass2}, \ref{fig:age1}, and \ref{fig:age2}). These quantities are typically estimated under standard assumptions, including the relation $L  4\pi R^2 \sigma T_{eff}^4$, which assumes a single blackbody approximation. Therefore, the reported values should be interpreted within the context of the physical assumptions of the adopted evolutionary models. The dispersion among models thus reflects an additional source of systematic uncertainty, particularly relevant for young, magnetically active stars.

\item For young stars, non-photospheric contributions such as accretion luminosity may affect the observed fluxes. In addition, photometric variability caused by variable accretion, surface spots, or circumstellar disc occultations can alter the inferred \teff and \logLs. These effects may shift the stellar position in the HRD, leading to systematic biases in the derived masses and ages. The reported parameters should therefore be interpreted with caution for strongly variable or actively accreting sources.

\item For higher-mass members approaching the zero-age main sequence (ZAMS), evolutionary tracks and isochrones become densely packed in the HRD, producing a strong age degeneracy. In this regime, small uncertainties in \teff and \logLs propagate into large age uncertainties, potentially biasing the inferred ages toward younger values. Consequently, age estimates for stars near the ZAMS should be regarded as intrinsically uncertain.

\item \massage assumes solar metallicity, and for PARSEC and MIST stellar evolution models, it covers ages up to 1 Gyr. However, the code can be easily adapted to other metallicities, different age ranges, and to include additional evolutionary models.
\end{itemize}

Despite these limitations, \massage provides a flexible and transparent framework for estimating stellar parameters, enabling users to evaluate how different observational uncertainties and model assumptions affect the inferred masses and ages.

Code availability. The Python implementation of \texttt{MassAge} will be made publicly available upon publication of this paper at
\url{https://github.com/jesush149/MassAge_py}

\section{The median SEDs of Taurus and the ONC subsample}
\label{app:medianSED}

We built the median SED of Taurus shown in Figure \ref{fig:medianONC} using the sample of \citet{Esplin2019} for K0-M6 stars, the corresponding median values listed  
in Table \ref{tab:median_Taurus_full}. The median and interquartile ranges were
normalized to the H band.
Table \ref{tab:median_Taurus_ONC_SpT} shows the Taurus median SEDs separated by spectral types bins, as shown in Figure \ref{fig:medianONC_sptyp}. The median SED for the ONC-Kim subsample, obtained by cross-matching our full CTTS sample with that of \citet{Kim2016}, is also given in Table \ref{tab:median_Taurus_ONC_SpT}.

\begin{table}
\centering
\caption{Taurus median H-normalized SED and quartiles for the \citet{Esplin2019} sample, built for K0--M6 stars}
\label{tab:median_Taurus_full}
\begin{tabular}{c c c c}
\hline
Wavelength [\micron] &
$\log_{10}\left[(\lambda F_\lambda)/(\lambda F_\lambda)_H\right]$ &
Lower & Upper \\
\hline
\hline
0.532  & -0.605 & -0.936 & -0.203 \\
0.673  & -0.293 & -0.478 & -0.009 \\
0.797  & -0.143 & -0.257 &  0.025 \\
1.235  &  0.047 &  0.002 &  0.085 \\
1.662  &  0.000 &  0.000 &  0.000 \\
2.159  & -0.146 & -0.174 & -0.084 \\
3.600  & -0.503 & -0.597 & -0.395 \\
4.500  & -0.646 & -0.749 & -0.476 \\
5.800  & -0.814 & -0.930 & -0.616 \\
8.000  & -0.897 & -1.068 & -0.660 \\
24.000 & -1.200 & -1.421 & -0.961 \\
\hline
\hline
\end{tabular}
\end{table}

\begin{table*}
\centering
\caption{Median H-normalized SEDs and quartiles by spectral-type range for Taurus, and for the ONC subsample cross-matched with the \citet{Kim2016} sample.}
\label{tab:median_Taurus_ONC_SpT}
\begin{tabular}{c c c c c c c}
\hline
 &
\multicolumn{3}{c}{Taurus} &
\multicolumn{3}{c}{ONC-Kim subsample} \\
\cline{2-4}
\cline{5-7}
Wavelength [\micron] &
Median & Lower & Upper &
Median & Lower & Upper \\
\hline
\hline

\multicolumn{7}{c}{\textbf{K0--K4}} \\
\hline
0.532  &  0.027 & -0.312 &  0.195 &  0.098 &  0.046 &  0.116 \\
0.673  &  0.108 & -0.239 &  0.188 &  0.138 &  0.095 &  0.303 \\
0.797  &  0.113 & -0.141 &  0.212 &  0.130 &  0.106 &  0.156 \\
1.235  &  0.063 &  0.022 &  0.120 &  0.060 &  0.018 &  0.075 \\
1.662  &  0.000 &  0.000 &  0.000 &  0.000 &  0.000 &  0.000 \\
2.159  & -0.116 & -0.170 & -0.036 & -0.139 & -0.174 & -0.096 \\
3.600  & -0.486 & -0.752 & -0.363 & -0.488 & -0.573 & -0.370 \\
4.500  & -0.666 & -0.853 & -0.455 & -0.587 & -0.717 & -0.530 \\
5.800  & -0.739 & -0.943 & -0.484 & -0.727 & -0.880 & -0.693 \\
8.000  & -0.857 & -1.183 & -0.610 & -0.814 & -0.969 & -0.641 \\
24.000 & -1.041 & -1.363 & -0.829 & -1.018 & -1.407 & -0.798 \\

\hline
\hline
\multicolumn{7}{c}{\textbf{K5--M2}} \\
\hline
0.532  & -0.301 & -0.597 & -0.025 & -0.358 & -0.461 & -0.173 \\
0.673  & -0.123 & -0.351 &  0.038 & -0.253 & -0.317 & -0.163 \\
0.797  & -0.030 & -0.185 &  0.080 & -0.077 & -0.141 & -0.033 \\
1.235  &  0.013 & -0.035 &  0.042 &  0.019 & -0.031 &  0.041 \\
1.662  &  0.000 &  0.000 &  0.000 &  0.000 &  0.000 &  0.000 \\
2.159  & -0.115 & -0.156 & -0.065 & -0.118 & -0.175 & -0.053 \\
3.600  & -0.462 & -0.551 & -0.326 & -0.498 & -0.571 & -0.290 \\
4.500  & -0.556 & -0.692 & -0.387 & -0.598 & -0.692 & -0.393 \\
5.800  & -0.685 & -0.851 & -0.508 & -0.734 & -0.853 & -0.582 \\
8.000  & -0.820 & -0.969 & -0.555 & -0.792 & -0.962 & -0.670 \\
24.000 & -1.114 & -1.280 & -0.929 & -0.932 & -1.101 & -0.673 \\

\hline
\hline
\multicolumn{7}{c}{\textbf{M3--M6}} \\
\hline
0.532  & -0.900 & -1.230 & -0.614 & -0.979 & -1.155 & -0.841 \\
0.673  & -0.436 & -0.573 & -0.233 & -0.588 & -0.762 & -0.510 \\
0.797  & -0.198 & -0.308 & -0.110 & -0.335 & -0.435 & -0.249 \\
1.235  &  0.081 &  0.047 &  0.097 &  0.013 &  0.001 &  0.057 \\
1.662  &  0.000 &  0.000 &  0.000 &  0.000 &  0.000 &  0.000 \\
2.159  & -0.166 & -0.189 & -0.138 & -0.132 & -0.153 & -0.121 \\
3.600  & -0.565 & -0.641 & -0.494 & -0.461 & -0.530 & -0.411 \\
4.500  & -0.700 & -0.806 & -0.597 & -0.607 & -0.680 & -0.502 \\
5.800  & -0.875 & -1.012 & -0.760 & -0.689 & -0.766 & -0.593 \\
8.000  & -0.992 & -1.139 & -0.837 & -0.613 & -0.811 & -0.580 \\
24.000 & -1.255 & -1.495 & -1.110 & -0.758 & -1.012 & -0.633 \\

\hline
\hline
\end{tabular}
\end{table*}


\bsp	
\label{lastpage}
\end{document}